\documentclass[fleqn,usenatbib]{mnras}

\usepackage{newtxtext,newtxmath}

\usepackage[T1]{fontenc}

\DeclareRobustCommand{\VAN}[3]{#2}
\let\VANthebibliography\thebibliography
\def\thebibliography{\DeclareRobustCommand{\VAN}[3]{##3}\VANthebibliography}

\usepackage{graphicx}	
\usepackage{amsmath}	
\usepackage{soul}
\usepackage{subfigure}
\usepackage{orcidlink}

\title[]{N$_2$H$^+$(1$-$0) as a tracer of dense gas in and between spiral arms}
\author[O. Feh\'er et al.]{
Orsolya Feh\'er$^{1}$\thanks{e-mail: fehero@cardiff.ac.uk}\orcidlink{0000-0002-0786-7307},
S. E. Ragan$^{1}$\orcidlink{0000-0003-4164-5588},
F. D. Priestley$^{1}$\orcidlink{0000-0002-5858-6265},
P. C. Clark$^{1}$\orcidlink{0000-0002-4834-043X},
and T. J. T. Moore$^{2}$
\\
$^{1}$School of Physics and Astronomy, Cardiff University, Queen's Buildings, The Parade, Cardiff C24 3AA, UK \\
$^{2}$Astrophysics Research Institute, Liverpool John Moores University, IC2, Liverpool Science Park, 146 Brownlow Hill, Liverpool L3 5RF, UK\\
}

\date{Accepted by MNRAS.}

\pubyear{2023}

\begin{document}
\label{firstpage}
\pagerange{\pageref{firstpage}--\pageref{lastpage}}
\maketitle

\begin{abstract}
Recent advances in identifying giant molecular filaments in galactic surveys allow us to study the interstellar material and its dense, potentially star forming phase on scales comparable to resolved extragalactic clouds. Two large filaments detected in the CHIMPS $^{13}$CO(3$-$2) survey, one in the Sagittarius-arm and one in an inter-arm region, were mapped with dense gas tracers inside a 0.06\,deg$^2$ area and with a spatial resolution of around 0.4 and 0.65\,pc at the distance of the targets using the IRAM 30\,m telescope, to investigate the environmental dependence of the dense gas fraction. The N$_2$H$^+$(1$-$0) transition, an excellent tracer of the dense gas, was detected in parsec-scale, elliptical clumps and with a filling factor of around 8.5\% in our maps. The N$_2$H$^+$-emitting areas appear to have higher dense gas fraction (e.g. the ratio of N$_2$H$^+$ and $^{13}$CO emission) in the inter-arm than in the arm which is opposite to the behaviour found by previous studies, using dust emission rather than N$_2$H$^+$ as a tracer of dense gas. However, the arm filament is brighter in $^{13}$CO and the infrared emission of dust, and the dense gas fraction determined as above is governed by the $^{13}$CO brightness. We caution that measurements regarding the distribution and fraction of dense gas on these scales may be influenced by many scale- and environment-dependent factors, as well as the chemistry and excitation of the particular tracers, then consider several scenarios that can reproduce the observed effect.
\end{abstract}




\section{Introduction}

\label{sec:intro}

Star formation is an essential element of all aspects and scales of astrophysics from the evolution of galaxies to the birth of planetary systems. The central aim of star formation research is to make a quantitative connection between the physical properties of interstellar gas and the star formation rate (SFR) within. The Kennicutt-Schmidt relation demonstrates that the surface densities of gas and star formation are related when averaged over a galaxy disc \citep{schmidt1959, kennicutt1998}, however, higher angular resolution surveys reveal significant galaxy-to-galaxy variations \citep{bigiel2008, shetty2014}. In nearby molecular clouds of the Milky Way the amount of star formation scales directly with the fraction of gas above a certain density threshold \citep{lada2010}, the so-called dense gas mass fraction but whether this diagnostic holds up in all Galactic environments is not well known. Whether a cloud resides in a spiral arm may be a key environmental factor \citep[e.g.][]{schinnerer2017}, but the current consensus is that the arms themselves do not significantly enhance star formation \citep{moore2012, eden2013, ragan2016}.

A major issue in interpreting the Kennicutt-Schmidt relation and its connection to similar Galactic results is insufficient knowledge of our tracers (density, gas temperature, radiation field, opacity and abundance conditions of their excitation) and the incongruity in their usage in extragalactic and galactic investigations. In extragalactic studies, $^{12}$CO is often used as a tracer of "dense" star forming molecular gas, while "diffuse" gas is probed by atomic hydrogen. In Galactic studies, CO serves as the "diffuse" tracer and dust or higher critical density molecular-line tracers (e.g. HCN, HCO$^+$, N$_2$H$^+$) probe the dense gas most closely associated with star formation. Among the possible dense gas tracers in the radio regime, N$_2$H$^+$ specifically was shown to be perhaps even better at tracing dense gas than HCN \citep{forbrich2014, shirley2015, kauffmann2017, pety2017, priestley2023a}. The molecule is linear and in a stable closed shell where all electrons are paired and there is only a weak diamagnetic field generated by rotation. The hyperfine structure (HFS) of the rotational transitions forms due to the interaction between this field and the electric quadrupole moments of the two nitrogens. There are seven HF components grouped in three, and their relative strength can be used to measure optical depth. N$_2$H$^+$ is more resistant to freeze-out than CO and an ideal line to trace the velocity structure of interstellar structures. N$_2$H$^+$ has been regularly observed in both low and high-mass Galactic star formation regions and protoplanetary disks \citep[e.g.][]{benson1994, turner1977, qi2013, fernandez2014, yu2018}. Its first extragalactic detection was by \citet{mauersberger1991} and since then it has been observed in a few nearby galaxies like M83 \citep{harada2019} and recently towards starburst galaxies as a part of the ALCHEMI project with ALMA \citep{martin2021}.

Existing observations of dense gas tracers in nearby galaxies have already given interesting results. A survey of HCN in normal and starburst galaxies by \citet{gao2004} revealed a tight correlation between infrared luminosity (a proxy for SFR) and HCN luminosity (a proxy for dense gas) when averaged over a galaxy, suggesting constant SFR per amount of dense gas. A targeted survey of dense clumps in the Milky Way found the same ratio on parsec scales \citep{wu2010}. However, another investigation of HCN from 300 dense clumps calculated an order of magnitude shortfall in the extrapolated total Galactic emission that would be needed for the Milky Way to satisfy the Gao-relation \citep{stephens2016}. High resolution studies of HCN and other dense gas tracers in nearby galaxies suggest that the IR-luminosity/HCN luminosity relation varies significantly with local environment \citep{bigiel2016, sun2018}. There must be other contributors to the overall HCN-luminosity budget of a galaxy and there might also be an intrinsic scatter from region to region.

Due to the angular resolution limits in studies of nearby galaxies, analogous studies of the Milky Way on parsec-scales can instead reveal the origin of dense gas tracer emission. A growing number of unbiased Galactic plane surveys have enabled us to begin linking small to large scales, thus probe the environmental dependence of the physical conditions of star formation. \citet{ragan2014} identified the first sample of velocity-coherent giant molecular filaments (GMFs) in the Galactic plane which are Milky Way-analogues of the scales probed in nearby galaxies. GMFs have been subsequently identified throughout the inner Galactic plane \citep[e.g.][]{zucker2015, abreu2016} and were found both along spiral arms and inside inter-arm regions. The two kinds of regions show differences in dense gas mass fraction and other properties, but due to limited resolution and sparse sampling of dense gas tracers \citep{wienen2012, shirley2013} we have a long way to go in acquiring a definitive view on how arm and inter-arm filaments differ. High-contrast, high-resolution, but lower critical density gas tracer measurements over significant sections of the Galactic plane like GRS \citep{jackson2006}, COHRS \citep{dempsey2013}, or CHIMPS \citep{rigby2016} will aid in constraining the details of spiral disc models, associate spectral features to star forming sites, and allow for source distances to be calculated, enabling the identification of more extragalactic-analogous structures. Two complexes in Orion \citep{pety2017, kauffmann2017} and the massive star-forming region W49 \citep{barnes2020} have been mapped in dense gas tracer emission as well, however, beside these targeted surveys of extreme star forming areas, sampling the full range of environmental conditions and the ordinary molecular gas that comprises most of the plane should also be held priority.

For this purpose, we used the unbiased, high angular resolution maps of $^{13}$CO(3$-$2) of the CHIMPS survey probing relatively high critical density material to select segments of large-scale filaments inside and between spiral arms for observations with the IRAM 30\,m telescope. We targeted dense gas tracers at 3\,mm and 1\,mm simultaneously, commonly used both in Galactic and extragalactic studies. In Section \ref{sec:data} we introduce the dataset and describe the data reduction methods. From Section \ref{sec:results} we focus on the tracer N$_2$H$^+$(1$-$0) and derive the spectral characteristics of the molecular emission for each mapped region. We derive the dense gas tracer-to-CO ratio through dense gas mass fraction-analogous parameters across each region and compare the results between arm and inter-arm structures. Using dendrogram-based clustering, we define N$_2$H$^+$ clusters and their characteristics, compare N$_2$H$^+$-bright and -faint areas and their dust and $^{13}$CO emission to find out more about our tracer and the environments it emits in. We discuss the results in Section \ref{sec:disc}, comparing them to those of other observed galactic clouds and the beam-averaged quantities found in nearby galaxies. We summarize our findings in Section \ref{sec:conc}.

\section{Data reduction and analysis methods} 
\label{sec:data}

\subsection{The CHIMPS survey}

The $^{13}$CO/C$^{18}$O(J\,=\,3$-$2) Heterodyne Inner Milky Way Plane Survey (CHIMPS) was carried out using the Heterodyne Array Receiver Program (HARP) on the 15\,m James Clerk Maxwell Telescope (JCMT) in Hawaii, covering approximately 18 square degrees in the region 27$\fdg$5$\lid l \lid$ 46$\fdg$4 and $|b|$ $\lid$ 0$\fdg$5. HARP is a 16-receptor focal-plane array receiver operating over the frequency range of 325$-$375\,GHz, and the Auto-Correlation Spectral Imaging System backend that was used in conjunction provided 250\,MHz bandwidth with 61.0\,kHz channel width. The velocity width per channel is 0.055\,km\,s$^{-1}$ and the observations were taken in a position-switching raster (on-the-fly) mode. The angular resolution of the observations is 15$\arcsec$. For more details on the observations, see \citet{rigby2016}.

The position-velocity diagrams for the two CO isotopologues from the CHIMPS survey integrated over the latitude axis show the spiral arms of the Milky Way clearly visible as continuous streams of emission, and inter-arm regions also appear as relatively emission-free areas separating the arms. The distribution of the gas fits reasonably well with a four-arm model with some deviations. Among these, a significant quantity of emission between the Scutum-Centaurus and Sagittarius arm was seen, more clearly than in previous surveys \citep[e.g.][]{dame2001, lee2001} or even in GRS or COHRS. The emission could be explained by a number of physical features: a minor spiral arm, an extension of the Scutum-Centaurus arm itself, a bridging or spur structure \citep{stark2006}, or a number of spurs forming coherent structures, extending for several degrees \citep{rigby2016}.

Specifically in the 36$\degr \lid l \lid$ 38$\degr$ longitude range of the CHIMPS survey area, two distinct $^{13}$CO velocity features were seen: emission associated with the Sagittarius spiral arm around 50\,km\,s$^{-1}$ (approximately at $d$\,=\,3\,kpc) and emission at 70\,km\,s$^{-1}$ (approximately at $d$\,=\,5\,kpc) when considering the four-arm model of \citet{taylor1993} updated in \citet{clemens2004}. The segments are over 2 degrees in longitude, roughly equivalent to 175\,pc and 100\,pc in filament length, respectively. We selected six sub-regions along these two filaments, focusing both on areas with strong $^{13}$CO emission where dense gas tracers are likely to be excited and at least equal area of low-level emission in order to probe a range of gas conditions in both spiral-arm and inter-arm regions.

\begin{figure*}
    \centering
    \includegraphics[width=\linewidth]{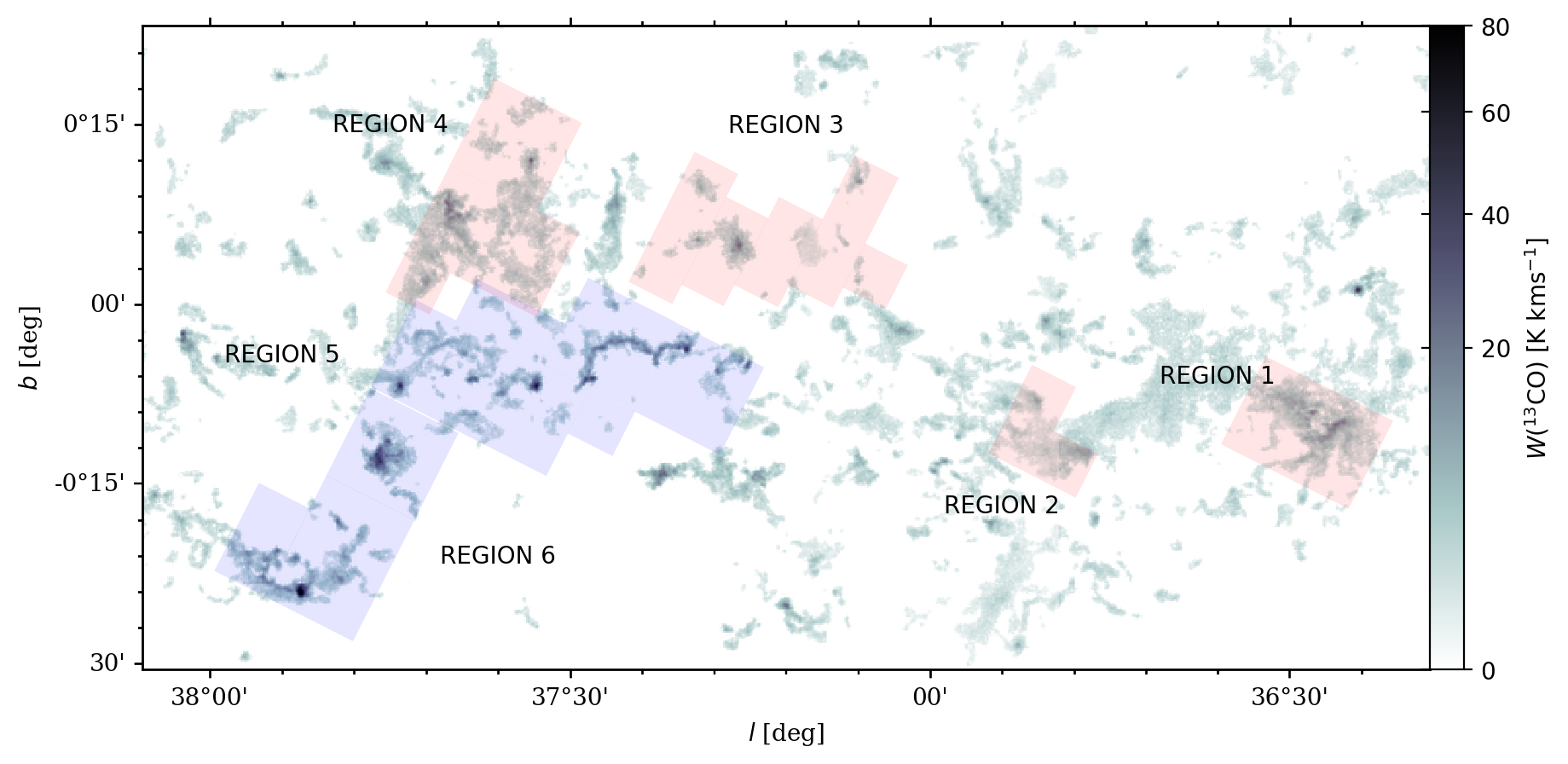}
    \caption{The six selected regions along arm (blue) and inter-arm filament segments (red) on the CHIMPS $^{13}$CO(3$-$2) integrated intensity maps. The colourmap is normalized with a square-root function to better visually represent the fainter $^{13}$CO structures.}
    \label{fig:chimpsfilaments}
\end{figure*}

\begin{figure}
\centering
    \includegraphics[width=\linewidth]{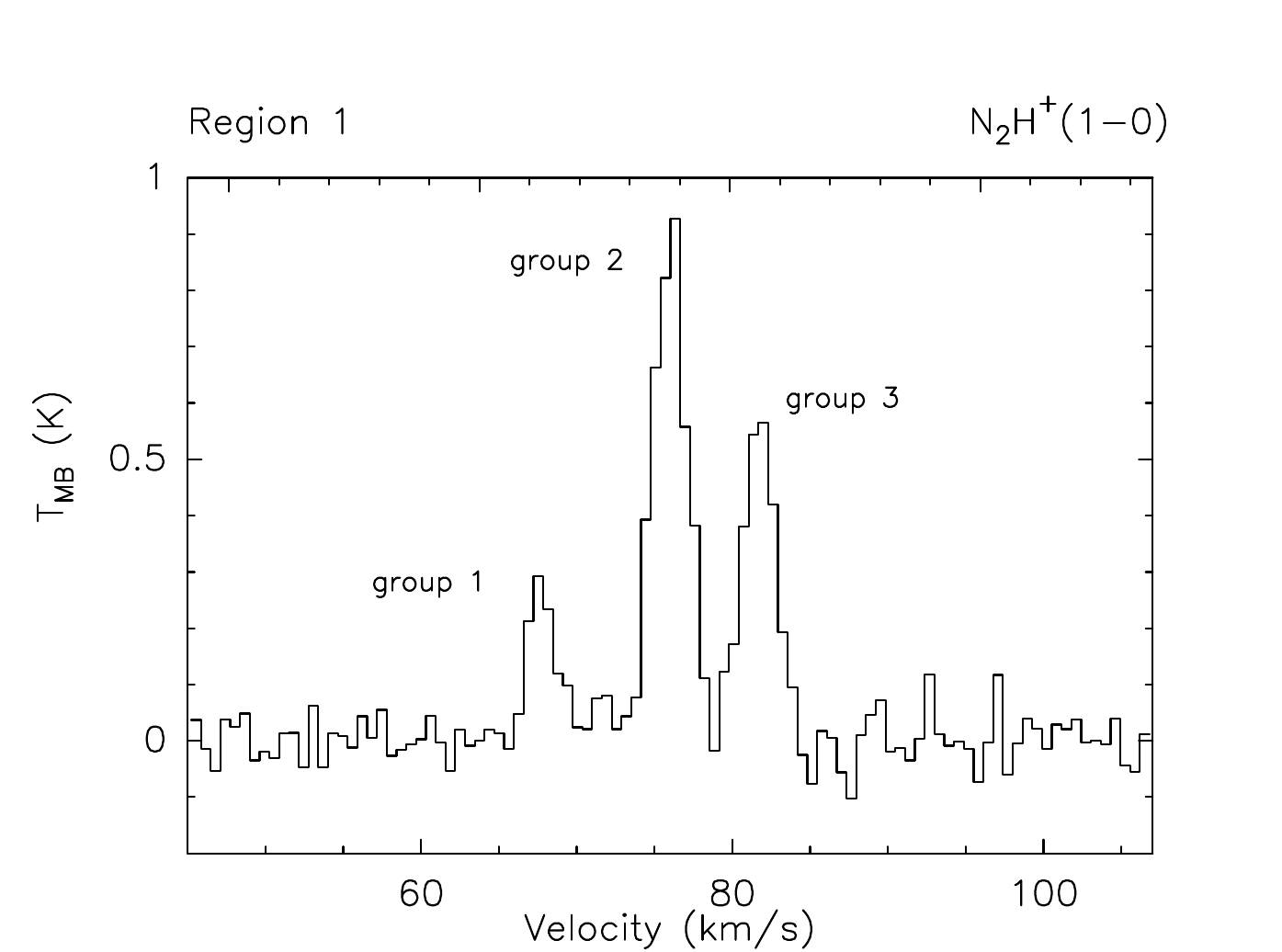}
    \caption{Example spectrum of the detected N$_2$H$^+$(1$-$0) transition in Region\,1, averaged in a beam at the position of the integrated intensity maximum. The hyperfine components are seen blended into three lines which are labeled group\,1$-$3.}
\label{fig:spectrum}
\end{figure}

\subsection{The IRAM 30\,m on-the-fly maps}
\label{data:iram}

The observations were performed with the IRAM 30\,m telescope under the project IDs 033-17 and E02-22 during 8$-$14\,August, 2017 and 22\,February, 2023. Six sub-regions were chosen towards the large-scale filamentary structures seen on the CHIMPS maps. The regions, four towards the inter-arm filament and two towards the Sagittarius arm filament, were mapped with the Eight Mixer Receiver (EMIR) at 1 and 3\,mm simultaneously. For the 3\,mm observations that appear in this study, the Fast Fourier Transform Spectrometer (FTS) was connected with the lower sideband tuned to 91.3\,GHz, providing 7.7\,GHz bandwidth and 200\,kHz spectral resolution covering the 85.8-93.4\,GHz range, resulting in a velocity resolution of around 0.6\,kms$^{-1}$ at the frequency of the N$_2$H$^+$(1$-$0) transition. During project 033-17 we obtained 120$\arcsec$\,$\times$\,120$\arcsec$ on-the-fly (OTF) maps with 12$\arcsec$ spacing using position-switching mode, observing 62 of these OTF maps to fill in the chosen filament regions. The half-power beam-width of the telescope at the tuning frequency was 26$\farcs$9. All in all, we spent 44 hours on the observations, reaching an average rms noise of 0.1\,K (in units of $T_{\rm A}$ antenna temperature) on each spectral position. Figure \ref{fig:chimpsfilaments} shows the selected regions against the integrated intensity of the CHIMPS $^{13}$CO(3$-$2) observations. All our targeted lines i.e. HCN(1$-$0), HCO$^+$(1$-$0), N$_2$H$^+$(1$-$0), C$^{18}$O(1$-$0) were detected towards both filamentary structures with good signal-to-noise ratio (S/N).

During the reduction and calibration of the data, the presence of negative emission (absorption structures) along the spectral axis made it obvious that the off-positions used in the position-switching observations were not entirely emission-clear. This was expected for CO isotopologues with lower critical densities but not for most dense gas tracers, however, the artifacts could be observed close to the lines in not only HCN and HCO$^+$, but even N$_2$H$^+$ spectra. To eliminate the effect, we requested time to observe the line emission on the off-positions of the OTF maps. During project E02-22, we performed pointed observations of 54 off-positions from the original measurement set contaminated by emission, using the same tuning setup and position-switching mode. The emission on the off-positions utilized this time were checked with pointed frequency-switched observations first. We reached an observing time of 4.4\,minutes per target position and an average rms noise of 0.02\,K (in units of $T_{\rm A}$) on each target position. In the end, to correct for the negative features on the OTF maps, we added the emission detected on the off-positions to each target spectra channel by channel. This cleared up the majority of affected positions except a single OTF-map in Region\,6 where strong negative emission of unknown origin (not detected when observing the off-position and not likely originating from any known strong radio source in the vicinity) appears on the spectra between 93.169$-$93.178\,GHz. Since this is very close to the science line, the area was excluded from further analysis.

In this paper, we use the observations of the N$_2$H$^+$(1$-$0) transition to trace the densest regions along the arm and inter-arm filaments. The seven HF components of the J\,=\,1 stage of N$_2$H$^+$(1$-$0) were found blended into three groups on our maps (as they generally do under many molecular cloud conditions) and occasionally blended further into a single line with multiple peaks. Figure \ref{fig:spectrum} presents a typical N$_2$H$^+$(1$-$0) spectrum. The data reduction of the spectra was done with the CLASS and GREG software packages of GILDAS. After baseline subtraction and extraction of the target lines on each spectral position, the OTF maps of a region were brought to the same absolute coordinate system, then the spectral map of the region was regridded using a convolution with a Gaussian and the recommended settings by CLASS to preserve Nyquist-sampling. The resulting cubes have a pixel size of 5$\farcs$55 while the native characteristics of the spectral axis were not changed. The antenna temperatures in the raw data files were converted to $T_{\rm MB}$ main beam brightness temperatures using the "modify beam eff /ruze" command in CLASS which uses the values stored for the 30\,m telescope to compute the correction as a function of frequency with Ruze's equation \citep{ruze1966} and rescales the spectra. This correction set the main beam efficiency to 0.81 for all the observed spectra on all channels of the extracted N$_2$H$^+$(1$-$0) line.

The integrated intensity, $W$(N$_2$H$^+$)\,=\,$\int T_{\rm MB}{\rm d}v$, of the line was calculated on each pixel by taking the 0th moment of the spectra (see Section \ref{n2hresults}). To derive well-constrained $v_{\rm LSR}$ central velocities and $\Delta v$ linewidths from the N$_2$H$^+$(1$-$0) spectra, we first fitted Gaussian-profiles to the main group (group\,2) of the line using the $gauss$ fitting procedure in CLASS on only those pixels where the S/N of $T_{\rm MB,2}$, the main beam brightness temperature of group\,2, was higher than 3. The average N$_2$H$^+$(1$-$0) spectral parameters detected in the regions are summarized in Table \ref{tab:regions_spectra}. We detect line strengths generally in the interval of 0.2$-$4\,K. The HFS linewidths are between 1$-$3\,km\,s$^{-1}$ with the highest values in Region\,6 and the lowest in the lower velocity component of Region\,1. We note that due to the generally low S/N spectral data on Region 2, results on this area need to be evaluated cautiously.

Following this, the \textit{hfs} method in CLASS was used to derive $\tau_{\rm g2}$, the optical depth of group 2, $v_{\rm LSR, HFS}$ HFS central velocity and $\Delta v_{\rm HFS}$ HFS linewidth with the limit of S/N\,$\gid$\,3 imposed onto the $T_{\rm MB,1}$ main beam brightness temperature of group\,1, additionally. On these spectral positions, $T_{\rm ex}$ excitation temperature, $\tau$ optical depth, and $N$(N$_2$H$^+$) N$_2$H$^+$ column density could also be calculated (see Appendix \ref{app:n2h} for the equations).

\begin{table}
    \centering
    \begin{tabular}{c|l|c|l|l|l|}
\multicolumn{1}{c|}{Region} & \multicolumn{1}{c|}{Filament} & \multicolumn{1}{c|}{C} & \multicolumn{1}{c}{$T_{\rm MB}$}& \multicolumn{1}{c|}{$v_{\rm LSR}$} & \multicolumn{1}{c|}{$\Delta v_{\rm HFS}$} \\
 & & & \multicolumn{1}{c}{[K]} & \multicolumn{1}{c|}{[km\,s$^{-1}$]} & \multicolumn{1}{c|}{[km\,s$^{-1}$]} \\
 \hline
 \hline
 1 & inter-arm & 1 & 0.2$-$1.2 & 77.8 (6.4) & 2.2 (1.8)  \\
   & &           2 & 0.2$-$0.6 & 52.3 (4.4) & 1.1 (0.01) \\
 2 & inter-arm & 1 & 0.4$-$1.2 & 78.9 (4.5) & 1.8 (1.8) \\
   & &           2 & 0.4$-$1.1 & 53.9 (3.2) & ...       \\
 3 & inter-arm & 1 & 0.2$-$3.7 & 89.9 (4.1) & 1.8 (0.8) \\
 4 & inter-arm & 1 & 0.2$-$2.8 & 84.6 (3.1) & 2.3 (1.8)  \\
 & &             2 & 0.2$-$1.2 & 42.6 (5.2) & 1.6 (1.1)  \\
 5 & arm & 1 & 0.2$-$2.8 & 51.1 (6.0) & 2.5 (3.5)   \\
 6 & arm & 1 & 0.2$-$3.5 & 65.1 (3.4) & 2.7 (1.3)   \\
 \hline
    \end{tabular}
    \caption{Average spectral parameters from the N$_2$H$^+$(1$-$0) emission on the observed regions. (1) Number of region on Fig. \ref{fig:chimpsfilaments}; (2) Galactic environment seen on CHIMPS maps; (3) Number of velocity component; (4) Observed line intensities; (5) Average $v_{\rm LSR}$ on the region from HFS fitting of N$_2$H$^+$(1$-$0) and its standard deviation; (6) Average $\Delta v_{\rm HFS}$ linewidth on the region from HFS fitting and its standard deviation. The averages were calculated on the spectral positions where the peak intensity of the N$_2$H$^+$ main group (group\,2) was detected with S/N\,$\geq$\,3 (see Section \ref{data:iram}).}
    \label{tab:regions_spectra}
\end{table}

\subsection{Clustering with the dendrogram method}
\label{sec:disc:ssec:dendro}

In order to better characterize the structures detected in N$_2$H$^+$(1$-$0) emission and to assess the condition of the emitting material, we applied dendogram clustering to our spectral cubes. There exists a variety of methods for extracting structures from 2D and 3D emission maps of molecular clouds. Dendrograms have the advantage of capturing the hierarchical nature of the interstellar medium, preserving the relationship between the various structures and providing a few easily adjustable parameters to define the depth of clustering. The analysis begins by locating emission peaks, then grouping the fainter surrounding pixels until two or more local maxima (so-called leaves) contain adjoining pixels. At this point, the two leaves merge at a branch. Fainter pixels are added to the merged structure until it is merged with another or reaches a user-defined noise threshold. At this point, it becomes a trunk. For more detailed discussion of the process, see e.g. \citet{rosolowsky2008}. 

Processing a molecular emission cube where HF components are present using dendrogram extraction produces spurious identifications at inaccurate source sizes and central velocities due to the split nature of the lines. To avoid this problem, we followed the example of \citet{keown2017} and constructed a simulated Gaussian emission data cube based on the emission of the main HF group in our N$_2$H$^+$(1$-$0) line emission. Similarly to their process involving NH$_3$(1,1) data, a Gaussian spectrum was fitted to the main HF group of the N$_2$H$^+$(1$-$0) line on each pixel over the observed maps, scaling the Gaussian functions to the main beam brightness temperature, Gaussian linewidth, and central velocity of the observed group\,2 on that position. Then random noise was added to the Gaussian spectrum with an rms equivalent to the average rms measured in the observed cube. \citet{keown2017} notes the warning of \citet{friesen2016} about error being introduced when using this method on spectra containing multiple velocity components. However, our data does not show any sign of close-by multiple velocity components except the two very well separated components seen towards the arm and inter-arm filaments, respectively.

We use the \textit{astrodendro} package to identify the hierarchical structures in the simulated Gaussian data cubes. The input parameters for the algorithm are: the minimum threshold value to consider in the cube (which we chose as two times the rms noise to be able to pick up fainter structures), the minimum difference in brightness between two structures before they are merged into a branch (which we also set as twice the rms noise) and the minimum number of pixels a structure must contain to remain independent (which we set as three times the pixel-per-beam value in the dataset). Using these parameters results in a fairly good selection of N$_2$H$^+$-emitting areas on our maps. However, in some cases, bad pixels on map or spectrum edges were picked up by the process, making later by-eye filtering necessary. We note that especially in the case of Region\,2 which has the lowest S/N among the observed spectral cubes, the locations of the identified clusters do not always coincide well with the local peaks on the N$_2$H$^+$(1$-$0) integrated intensity map.

We name the derived N$_2$H$^+$ clusters as: the letter R signifying "region", followed by the number of the region they appear in, then the letters CL signifying "cluster", followed by a number uniquely identifying the cluster on a certain region in an increasing manner e.g. R1CL1, R1CL2, etc. All in all, 40 N$_2$H$^+$ clusters were found by \textit{astrodendro}, 16 in the arm and 24 in the inter-arm filament. None of the extracted clusters show sub-structure and most of them are small, near-elliptical clumps with major axes generally smaller than 0$\farcm$5, thus, parsec- and sub-parsec scale on the sky. They are approximated by the dendrogram extraction process with ellipses.

\begin{figure*}
\centering
    \subfigure{\includegraphics[width=0.44\linewidth]{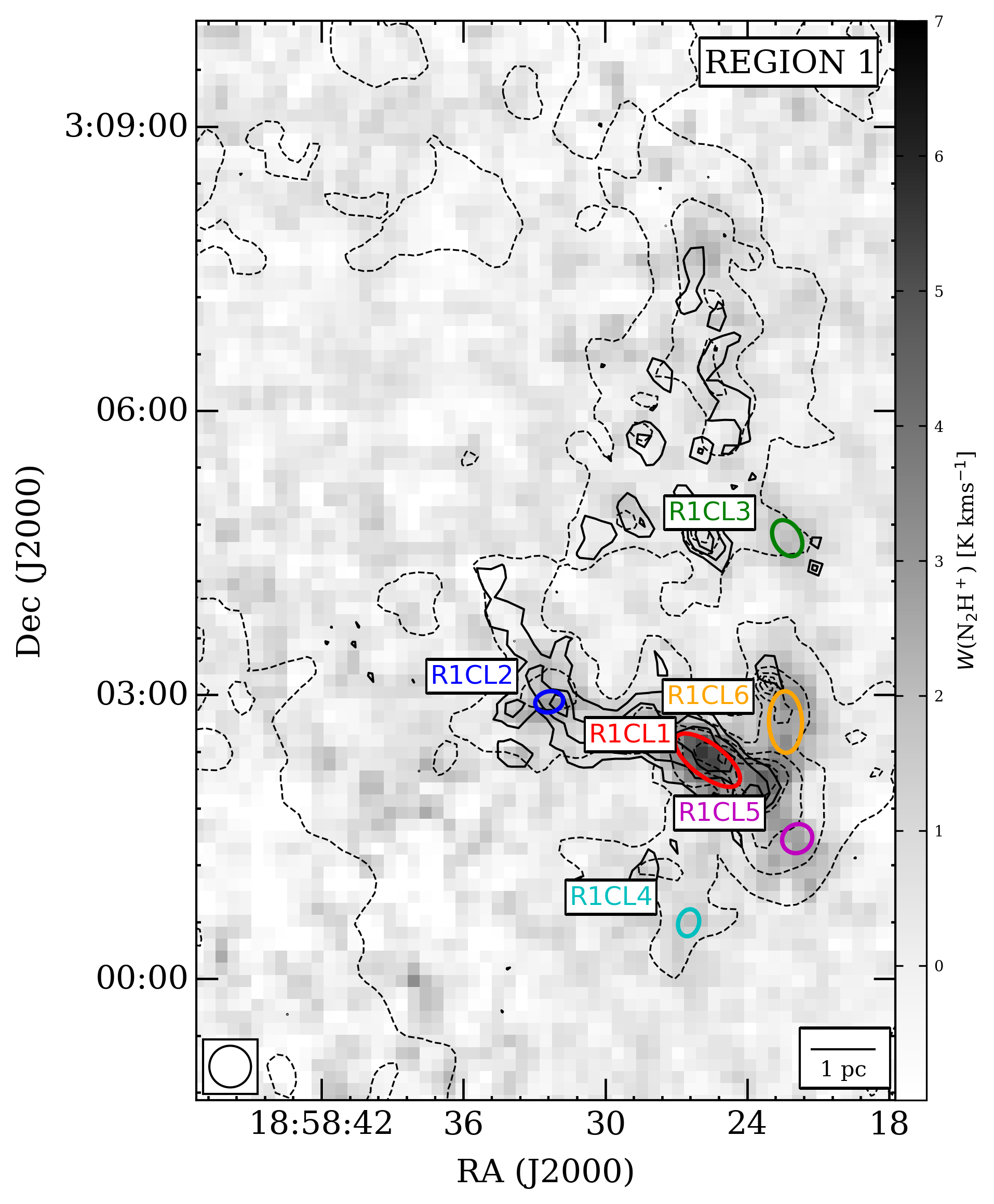}}
    \subfigure{\includegraphics[width=0.44\linewidth]{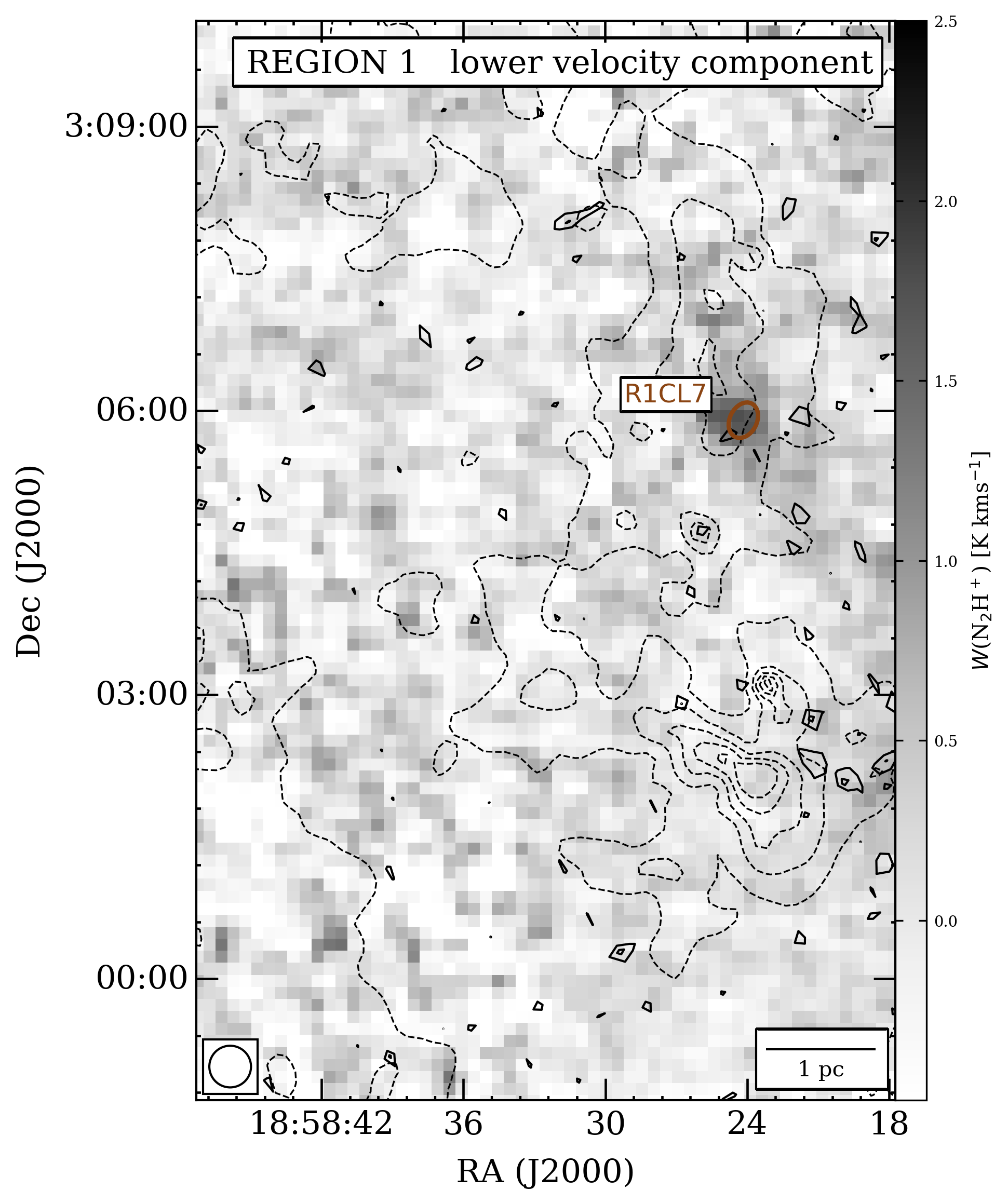}}
    \subfigure{\includegraphics[width=0.45\linewidth]{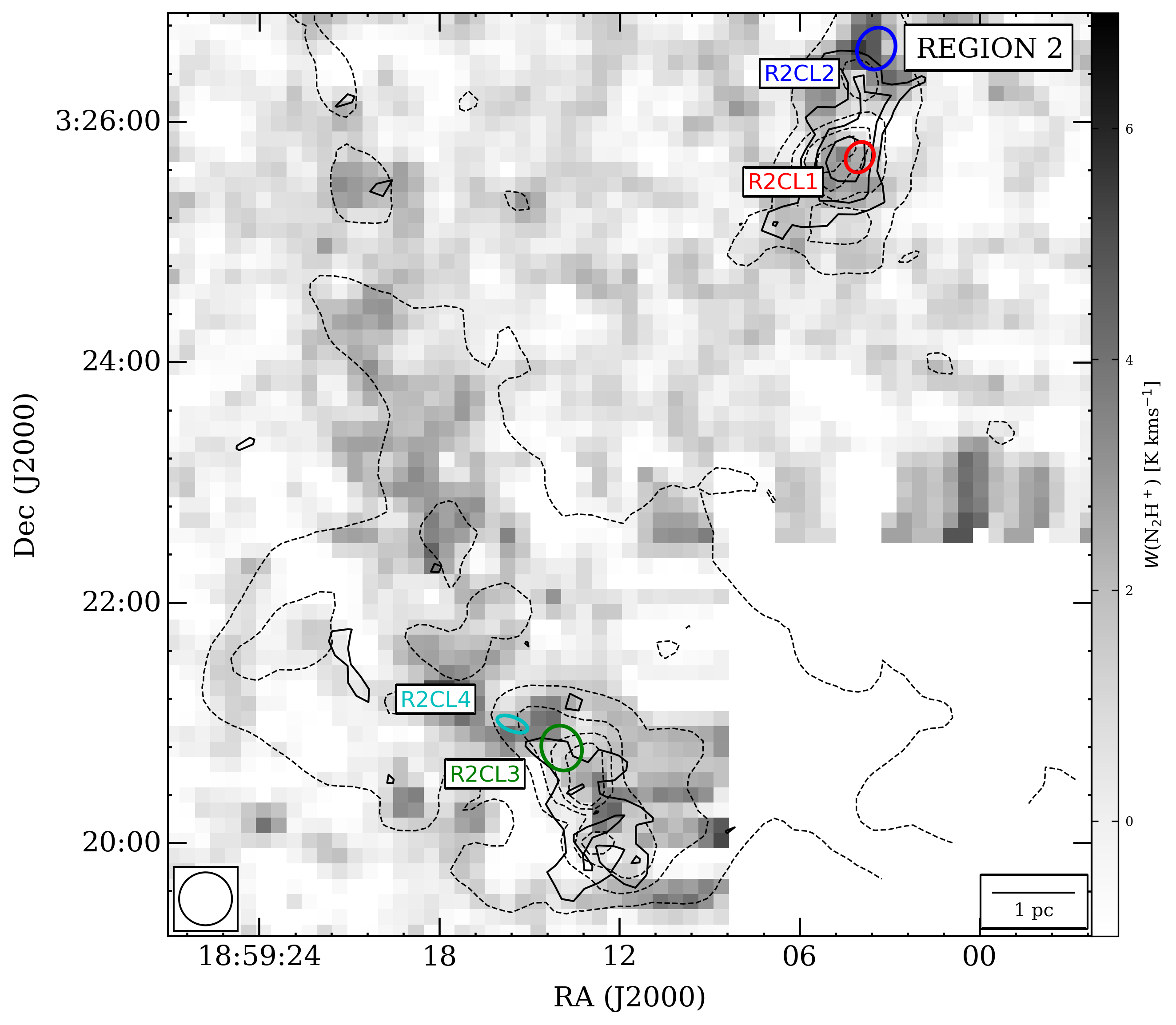}}
    \subfigure{\includegraphics[width=0.45\linewidth]{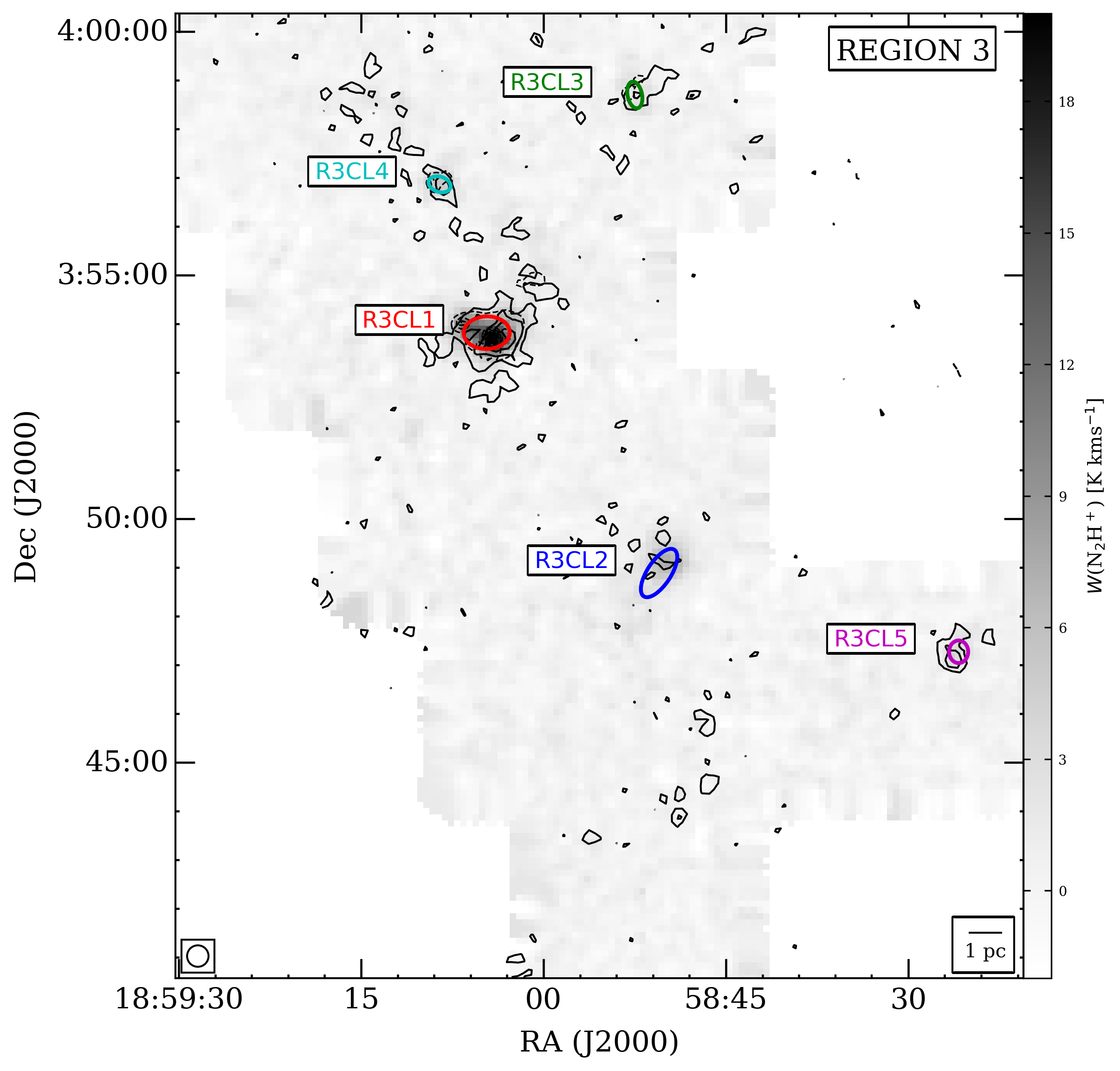}}
    \subfigure{\includegraphics[width=0.45\linewidth]{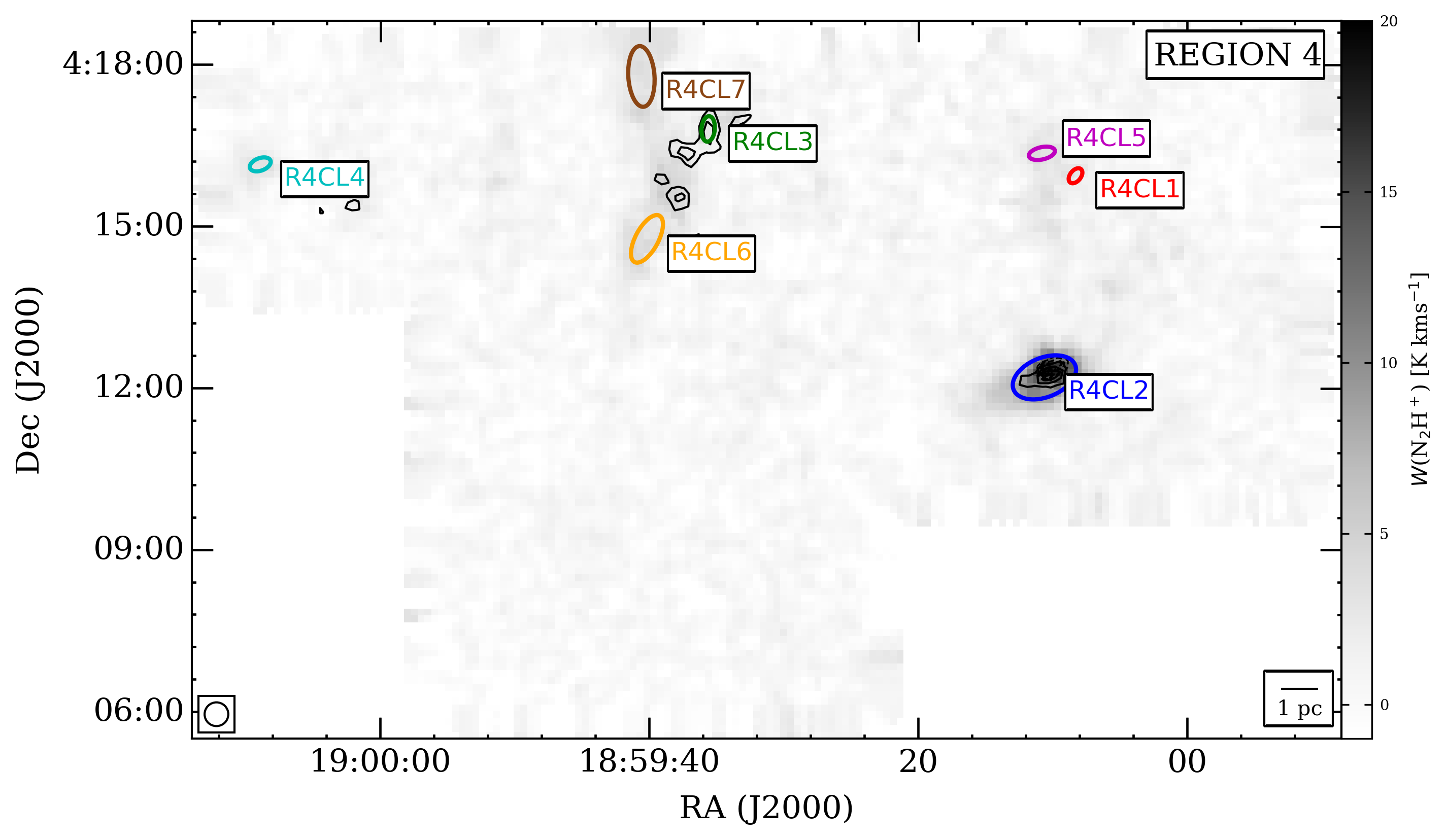}}
    \subfigure{\includegraphics[width=0.45\linewidth]{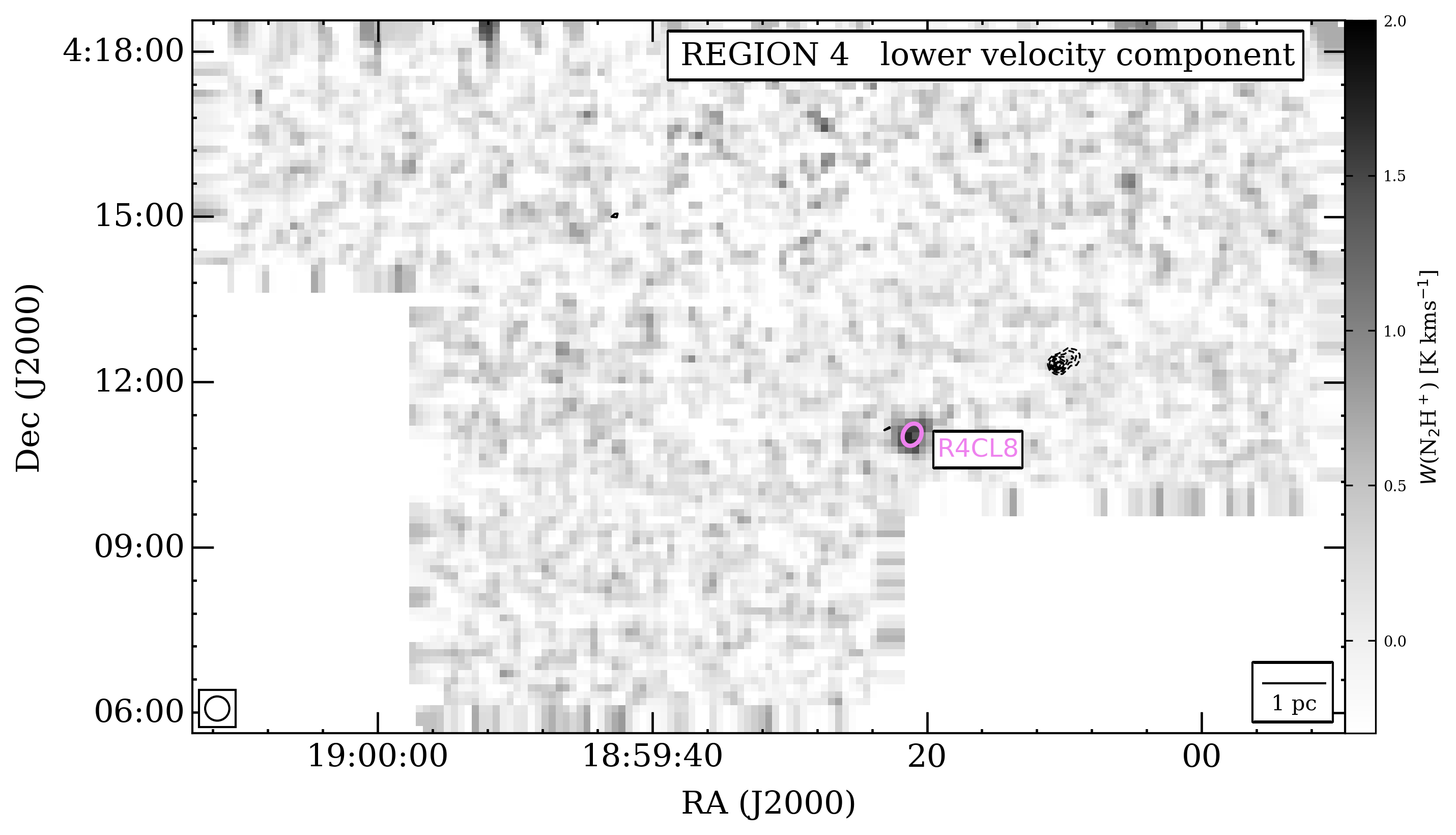}}
    \caption{Integrated intensity map of N$_2$H$^+$(1$-$0) for the four regions in the inter-arm filament, along with their lower velocity line component emission (for the integration limits see Section \ref{n2hresults}). Black contours mark the integrated intensity of $^{13}$CO(3$-$2) for the same velocity intervals where contour levels are at 40, 60, 80\% of the maxima of the maps (20 (10.2 on the lower velocity map), 16.9, 25.7, and 33.2 (20.8 on the lower velocity map) K\,kms$^{-1}$, respectively). The dashed black contours mark $N$(H$_2$)$_{\rm dust}$ where contour levels are are at 20, 30...80\% of the maximum (3.1, 2.7, 18.5, and 26.0\,$\times$\,10$^{22}$\,cm$^{-2}$, respectively). The coloured ellipses mark the N$_2$H$^+$ clusters extracted with the dendrogram clustering.}
    \label{fig:region1_mapcl}
\end{figure*}

\begin{figure*}
\centering
    \subfigure{\includegraphics[width=0.49\linewidth]{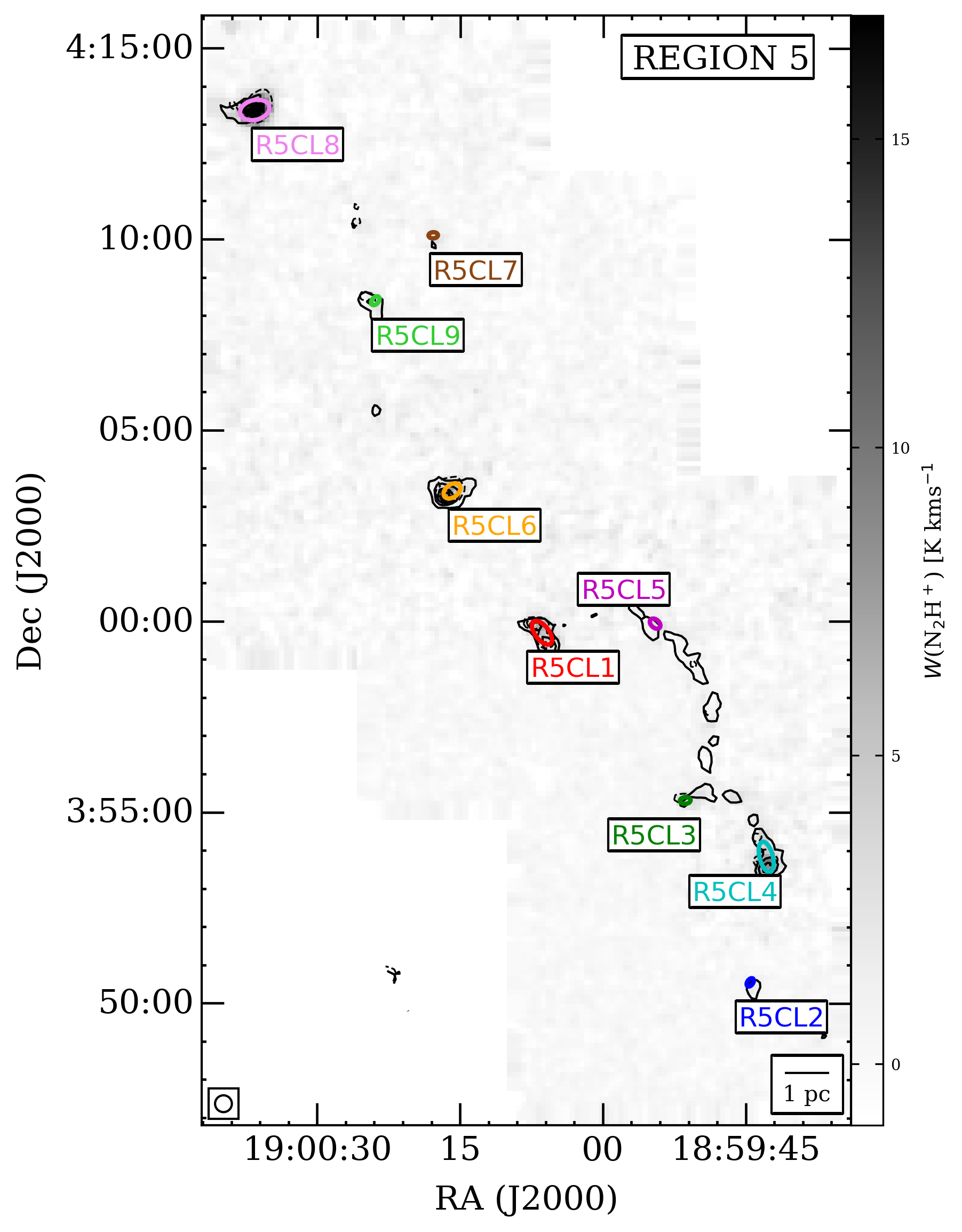}}
    \subfigure{\includegraphics[width=0.49\linewidth]{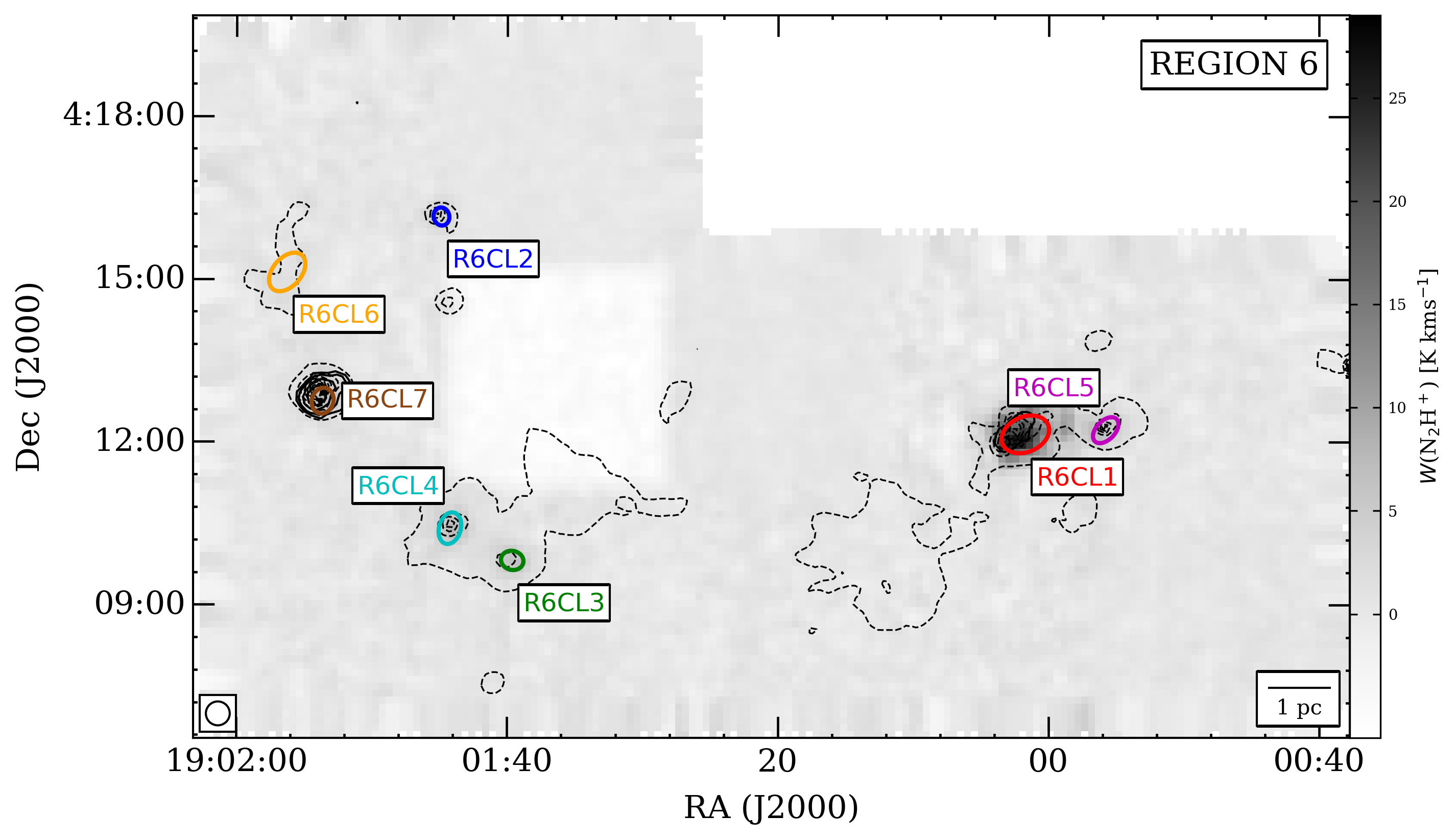}}
    \caption{Integrated intensity map of N$_2$H$^+$(1$-$0) for the two regions in the inter-arm filament (for the integration limits see Section \ref{n2hresults}). Black contours mark the integrated intensity of $^{13}$CO(3$-$2) for the same velocity intervals where contour levels are at 40, 60, 80\% of the maxima of the maps (70.9 and 112.0\,K\,kms$^{-1}$, respectively). The dashed black contours mark $N$(H$_2$)$_{\rm dust}$ where contour levels are are at 10, 20...80\% of the maximum (25.6 and 19.6\,$\times$\,10$^{22}$\,cm$^{-2}$, respectively). The coloured ellipses mark the N$_2$H$^+$ clusters extracted with the dendrogram clustering.}
    \label{fig:region5_mapcl}
\end{figure*}

\subsection{Hi-GAL continuum maps and PPMAP}

The Herschel Infrared Galactic Plane Survey (Hi-GAL) mapped the inner Milky Way in five infrared bands (70, 160, 250, 350 and 500 $\micron$) to encompass the peak of the spectral energy distribution of cold dust for 8\,$\lid$\,$T_{\rm dust}$\,$\lid$\,50\,K. In our current study, we made use of the $T_{\rm dust}$ and $N$(H$_2$)$_{\rm dust}$ maps obtained for the entire Hi-GAL survey range by \citet{marsh2017} using the PPMAP procedure \citep{marsh2015}.

The PPMAP tool is a step beyond conventional approaches of calculating column densities in that it does not assume that $T_{\rm dust}$ is uniform in the line of sight. It also enables achieving higher spatial resolutions, since smoothing of the input continuum maps is not required; instead, PPMAP uses the point spread functions of the telescopes directly. Given a set of observational images of dust emission at multiple wavelengths, the associated point-spread functions, a dust opacity law, a grid of temperature values, and assuming that the dust is optically thin, PPMAP performs a non-hierarchical Bayesian process to generate a density function equivalent to a differential column density as a function of angular position and dust temperature.

For the processing of the Hi-GAL data, \citet{marsh2017} used a power-law form for the dust opacity law:

\begin{equation}
\kappa(\lambda)=0.1\,\mathrm{cm^2\,g^{-1}}\left(\frac{\lambda}{300\micron}\right)^{\beta}
\end{equation}

\noindent with the spectral index $\beta$\,=\,2 based on previous studies. According to \citet{juvela2012} and \citet{juvela2015} the value of $\beta$ varies between 1.8 and 2.2 in the cold dense interstellar medium, anti-correlates with temperature, and correlates with column density and Galactic position. Using $\beta$\,=\,2 with \textit{Herschel} measurements is appropriate for dense clumps which were expected to be found by the Hi-GAL survey and which are assumed by us to be traced by N$_2$H$^+$(1$-$0) emission as well. The reference opacity 0.1\,cm$^2$g$^{-1}$ at 300\,$\micron$ is defined with respect to total mass (dust plus gas) and is consistent with a gas to dust ratio of 100 \citep{hildebrand1983}. 

The output of PPMAP with these parameters is firstly, an image cube of differential column densities with 6$\arcsec$ pixels and 12$\arcsec$ spatial resolution, and 12 values along the temperature axis, covering the range 8$-$50\,K in $T_{\rm dust}$. The unit of differential column density thus is H$_2$ molecules per square centimetre per degree Kelvin. The output also includes a corresponding image cube of uncertainties and two more maps: total column density $N$(H$_2$)$_{\rm dust}$ and density-weighted mean dust temperature, $T_{\rm dust}$. We use these last two maps in this paper. For more information on PPMAP and its use in processing the Hi-GAL data, see further explanations by \citet{marsh2015} and \citet{marsh2017}.

A catalogue containing the physical properties of compact sources was also released by \citet{elia2021} based on the Hi-GAL dataset covering the full Galactic plane with more than 150,000 entries, presenting flux density, distance, size, mass, temperature, surface density, luminosity, and classification into evolutionary categories. We use this data to further characterize the gas and the star formation in our target regions.

\subsection{Measuring the dense gas fraction}
\label{sec:dgmf}

To characterise the volume of gas most likely related to star formation and assess its distribution, environmental dependence, and relation to observable star formation activity tracers, we calculate the dense gas mass fraction-analogous quantities $f'_{\rm DG}$ and $h_{\rm N_2H^+}$.

We define

\begin{equation}
    f'_{\rm DG}=\frac{W(\mathrm{N_2H^+})}{W\mathrm{(^{13}CO)}}
\end{equation}

\noindent after the example of \citet{lada2012} and  more recently \citet{bigiel2016} who examined the variation of dense gas in M51 by showing the trends of the HCN-to-CO ratio across the Galaxy disk and defined $f_{\rm DG}$\,=\,$M_{\rm dense}$/M$_{\rm gas}$. They use $^{12}$CO to trace low density gas with volume densities of $n \gid$ 10$^2$\,cm$^{-3}$ and HCN with the critical density at the order of 10$^6$\,cm$^{-3}$ to trace the dense gas. We note however that \citet{kauffmann2017} argue that critical densities do not solely control how line emission couples to dense gas and e.g. HCN already traces characteristic densities at the order of 10$^3$\,cm$^{-3}$ \citep{gerwyn2023}. In contrast, as discussed before, they regard N$_2$H$^+$ as the only molecular species clearly selectively connected to dense gas, a result that is supported by recent theoretical work \citep{priestley2023a, priestley2023b}. 

The usage of $^{13}$CO(3$-$2) in place of $^{12}$CO(1$-$0) is a similarly complex issue. The optically thin critical density of $^{12}$CO(1$-$0) is 10$^3$\,cm$^{-3}$ at 10\,K, however, the transition is almost always strongly optically thick in molecular clouds, so due to radiative trapping, strong lines may be observed in gas with lower than 10$^2$\,cm$^{-3}$ effective critical densities \citep{yancy2015}. The isotopologue $^{13}$CO generally traces similar densities around 10$^3$\,cm$^{-3}$, however, its optical depth rarely increases above $\tau$\,=\,1$-$3. Thus the radiative trapping effects are not so important and $^{13}$CO will generally trace the somewhat denser parts of a molecular structure. Critical density is also not only the function of species but transition, with higher transitions tracing higher densities. The (3$-$2) transition of $^{13}$CO has a critical density of 1.6\,$\times$\,10$^4$\,cm$^{-3}$ as noted by \citet{rigby2016} based on \citet{schoier2005} thus it is more sensitive to higher density gas. The amount of detectable emission under the critical density is dependent on the density and the excitation temperature, and we expect to not trace some of the diffuse molecular material on the edges of the clouds using the CHIMPS $^{13}$CO(3$-$2). One advantage however, is that confusion with separate structures at different distances along the line of sight that the (1$-$0) transition might suffer from is minimized. With these considerations we use the above defined $f'_{\rm DG}$, the ratio of the integrated intensities of $^{13}$CO(3$-$2) and N$_2$H$^+$(1$-$0), as analogous to the dense gas mass fraction.

The mosaic $^{13}$CO(3$-$2) cube corresponding to our target area on the sky from the CHIMPS survey is available publicly in the CANFAR archive\footnote{\url{http://dx.doi.org/10.11570/16.0001}}. We integrated the $^{13}$CO(3$-$2) emission in the same velocity intervals that we detected N$_2$H$^+$(1$-$0) at and used these respective maps when computing $f'_{\rm DG}$ or comparing the integrated intensities of $^{13}$CO and N$_2$H$^+$, to make sure we are comparing radiation emitted from the same volume of material (see Section \ref{n2hresults}). The N$_2$H$^+$(1$-$0) maps with originally 5$\farcs$5 pixel size were resampled to the CHIMPS pixel size (7$\farcs$6) for these purposes. To be able to examine tendencies, we calculated $f'_{\rm DG}$ in each beam in our target regions using the beam-size of the IRAM telescope (26$\farcs$9) which is larger than the CHIMPS beam (15$\farcs$2).

We can also define

\begin{equation}
    h_{\rm N_2H^+}=\frac{W(\mathrm{N_2H^+})}{N(\mathrm{H_2})_{\rm dust}}.
\end{equation}

\noindent following the analysis of \citet{kauffmann2017} who used the line-to-mass ratio $h_{\rm Q}$\,=\,$W$(Q)/A$_{\rm V}$ in their analysis. Since $A_{\rm V}\,\propto$\,$N$(H$_2$), $h_{\rm Q}$ essentially measures the intensity of line emission per H$_2$ molecule. The trend of $h_{\rm Q}$ versus $N$(H$_2$) is non-trivial and differs between molecules. We resampled the $W$(N$_2$H$^+$) maps from their 5$\farcs$5 pixels to 6$\arcsec$ which is the Hi-GAL pixel size, to be able to compare the two, and when computing the beam-averaged values of $h_{\rm N_2H^+}$, we used the beam-size of the IRAM telescope, 26$\farcs$9, since it is larger than the spatial resolution of the PPMAP output $N$(H$_2$)$_{\rm dust}$ map, 12$\arcsec$.

Both $f'_{\rm DG}$ and $h_{\rm N_2H^+}$ can be thought of as a measure of the dense gas fraction. We emphasize here that most previous studies use the parameter "dense gas mass fraction", usually calculating the masses of total gas and dense gas inside a structure and comparing the results, while the parameters calculated in the present work are not mass fractions, thus we use "dense gas fraction" when talking about $f_{\rm DG}$ and $h_{\rm N_2H^+}$.

The used parameters $f_{\rm DG}$ and $h_{\rm N_2H^+}$ have their own caveats. CO becomes optically thick then starts depleting at high volume densities and we also have increasing evidence that a large amount of material exists as CO-dark gas in giant molecular clouds \citep{wolfire2010, glover2011, rowan2014}. Thus $^{13}$CO might underestimate the total gas mass in varying degrees in different environments. $N$(H$_2$)$_{\rm dust}$ calculated from far-infrared or sub-millimeter continuum radiation measures the total amount of material in a column at each line-of-sight, however, without velocity information, background and foreground clouds not associated with the target object may also contribute to the computed quantity. Biases in determining $T_{\rm dust}$, using the incorrect dust spectral index, and erroneous assumption of the gas-to-dust ratio can introduce uncertainties in the measure as well. There is also a dependence of the N$_2$H$^+$ emission efficiency on $T_{\rm dust}$ as seen by \citep{barnes2020} and it shows a differing correlation with CO and HCN emission on different scales \citep{jimenez2023}. In spite of this, recent advances in observing N$_2$H$^+$ in varying Galactic and extragalactic environments shed light to its unique chemistry allowing it to trace dense cores e.g. \citet{kauffmann2017, pety2017}. Examining its relation to $^{13}$CO is a satisfying measure of the dense gas fraction to first order and gives an excellent opportunity to further investigate the emission of this tracer, filling the scale-gap between its detections in Galactic clouds and in large extragalactic environments.

\section{Results} 
\label{sec:results}

\subsection{N$_2$H$^+$(1$-$0) in the arm and the inter-arm}
\label{n2hresults}

We detected the N$_2$H$^+$(1$-$0) transition towards all our target regions. We compute the filling factor as the ratio of pixels on which the main group of the N$_2$H$^+$(1$-$0) HFS shows S/N\,$>$\,3 and the total number of pixels. The filling factor is 8.6\% on the overall observed area (and 7, 8, 15, 10, 5 and 7\% in Regions\,1$-$6, respectively). We generally detect the seven HF components of the line blended into 3 groups as mentioned before, or even blended into a single line with multiple peaks. In lower S/N areas, occasionally the weakest group\,1 of the HFS was not detected. 

The $v_{\rm LSR}$ values of the detected N$_2$H$^+$(1$-$0) lines in the inter-arm filament regions (Regions\,1$-$4) correspond well with the 60$-$105\,km\,s$^{-1}$ velocity range defined from the CHIMPS $^{13}$CO structure there, but in smaller areas, another, secondary velocity components in the 26$-$52 and 42$-$61\,km\,s$^{-1}$ range can also be seen. Some of these were later identified as clusters R1CL7 and R4CL8. Towards Region\,5 and 6, the two arm filament sections, the main source of the N$_2$H$^+$(1$-$0) emission appears between 31$-$64\,km\,s$^{-1}$ and 50$-$77\,km\,s$^{-1}$, respectively. These are the velocities where we expect structures connected to the Sagittarius arm to lie. We only find localized emission at differing velocities on these two regions: weak emission in Region\,5 between 6.8$-$32.5\,kms$^{-1}$ and in Region\,6 between 69$-$94\,kms$^{-1}$. The low velocity component in Region\,5 was excluded from further analysis because it falls outside the velocity region where the two target filaments were defined on the CHIMPS maps. The high-velocity emitting area in Region\,6 was excluded because it is close to the edge of the mapped area and the emitting structure likely extends to the area not covered by observations. Figure \ref{sag_allspectra} shows the spectra of all the distinct N$_2$H$^+$(1$-$0) velocity components observed in our regions.

\begin{figure*}
\centering
    \subfigure{\includegraphics[width=0.9\linewidth]{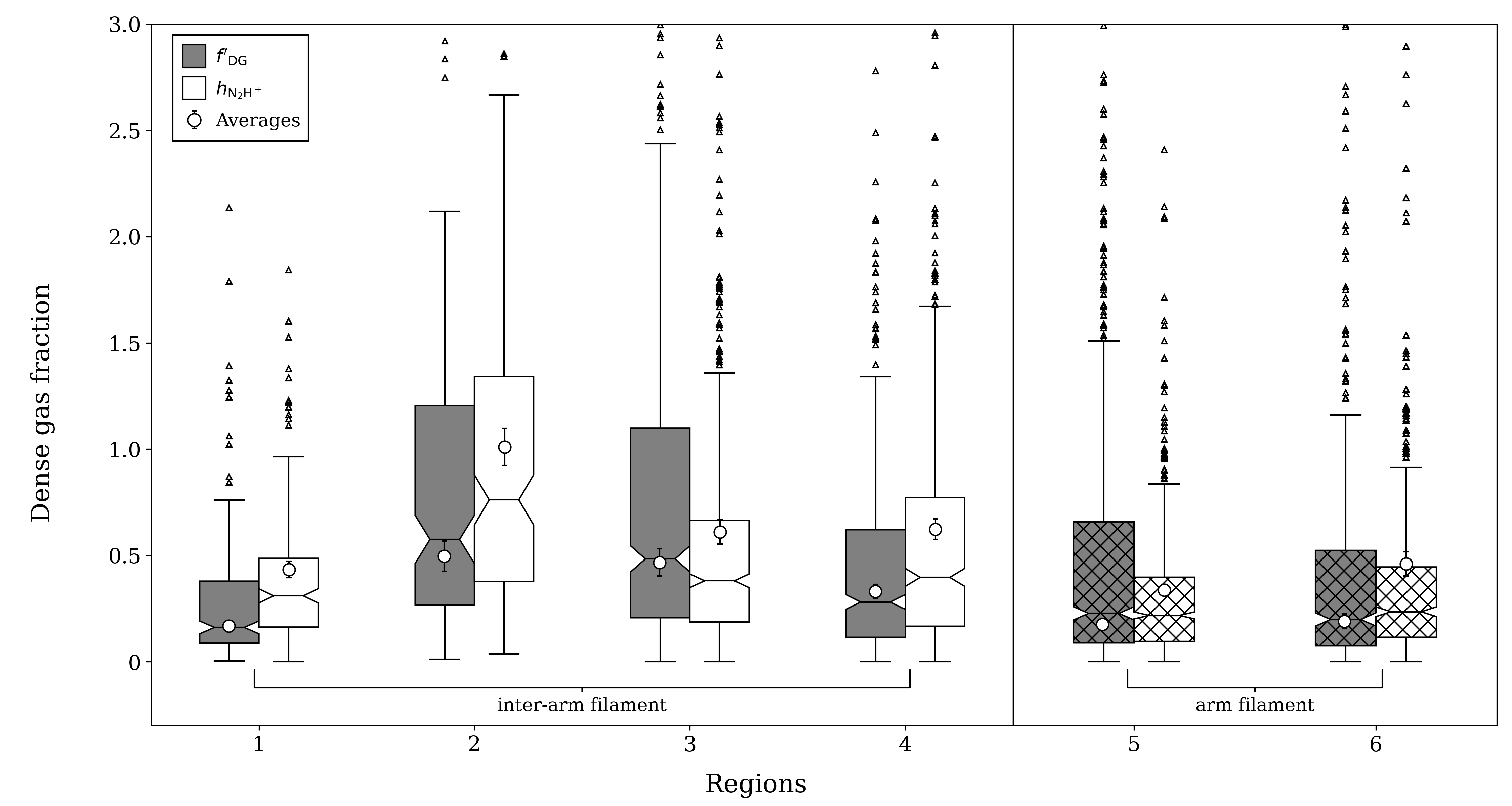}}
    \subfigure{\includegraphics[width=0.92\linewidth]{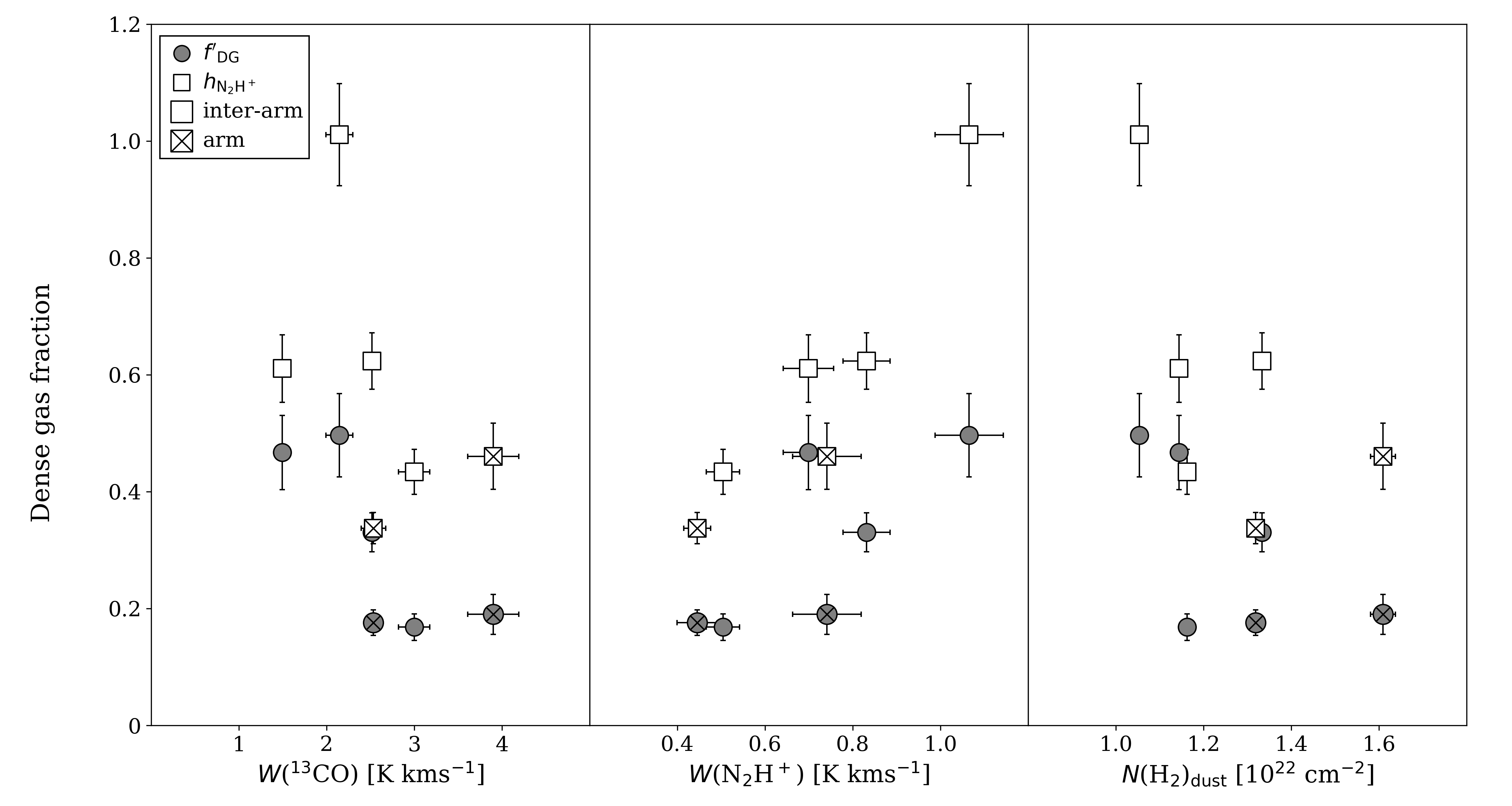}}
\caption{a) The distribution of the beam-averaged $f'_{\rm DG}$ and $h_{\rm N_2H^+}$ values (the latter in units of K\,kms$^{-1}$/10$^{22}$\,cm$^{-2}$) in the six regions. The boxes extend from the first quartile to the third quartile of the data, the notch indicates the confidence interval around the median (horizontal line), the whiskers extend to the farthest data point lying within 1.5\,$\times$ the inter-quartile range, and outlier points are indicated with triangles. The first, darker box marks the distribution of $f'_{\rm DG}$, the second, lighter box the distribution of $h_{\rm N_2H^+}$. Circle symbol marks the average values. Hashed boxes indicate the arm filament regions while non-hashed boxes the inter-arm filament regions. b) The average $f'_{\rm DG}$ and $h_{\rm N_2H^+}$ values in each region (the latter in units of K\,kms$^{-1}$/10$^{22}$\,cm$^{-2}$) versus the average $^{13}$CO(3$-$2) and N$_2$H$^+$(1$-$0) integrated intensities and the $N$(H$_2$)$_{\rm dust}$ of a region. The errors are computed as the error of the mean value in every case.}
\label{fig:allregion_fh}
\end{figure*}

Figures \ref{fig:region1_mapcl}-\ref{fig:region5_mapcl} show the N$_2$H$^+$(1$-$0) integrated intensity maps for the regions. As limits of the integration, we used the velocity ranges in which the N$_2$H$^+$(1$-$0) emission was detected in the different regions. For Region\,1$-$3, the higher velocity components were integrated uniformly between 60$-$105\,km\,s$^{-1}$ while the lower (except Region\,3 where there was no lower component found) between 42$-$61\,km\,s$^{-1}$. For Region\,4, the higher velocity component was integrated in the same interval as for the other inter-arm regions, 60$-$105\,km\,s$^{-1}$, however, the lower component was found and integrated between 26$-$52\,km\,s$^{-1}$. For Region\,5 and 6 the integration intervals were 31$-$64 and 50$-$77\,km\,s$^{-1}$. We note that the same velocity masks were used when computing the integrated intensities of the $^{13}$CO(3$-$2) line, since its emission was generally observed in the same velocity ranges as the N$_2$H$^+$(1$-$0) line, thus allowing us to look at the same volume of material without losing flux or S/N for either species.

The detected N$_2$H$^+$(1$-$0) emission roughly follows the $^{13}$CO(3$-$2) structures on the CHIMPS maps and the far-infrared dust emission on the Hi-GAL maps. However, in some cases, only weak $^{13}$CO(3$-$2) emission is detected on areas with strong N$_2$H$^+$(1$-$0) lines e.g. on Region\,3 the area identified as cluster R3CL2 or on Region\,6 the areas identified as clusters R6CL3 and R6CL4. The opposite can also be observed e.g. the long arc of $^{13}$CO(3$-$2) emission on Region\,5 shows very weak N$_2$H$^+$(1$-$0) lines. When looking at the integrated intensity maps of the secondary velocity components, there is generally little correlation between $W$(N$_2$H$^+$) and dust structures e.g. the lower velocity component map of Region\,4 in Figure \ref{fig:region1_mapcl}.

\begin{figure*}
\centering
    \includegraphics[width=0.9\linewidth, trim={0 2cm 0 2cm}, clip]{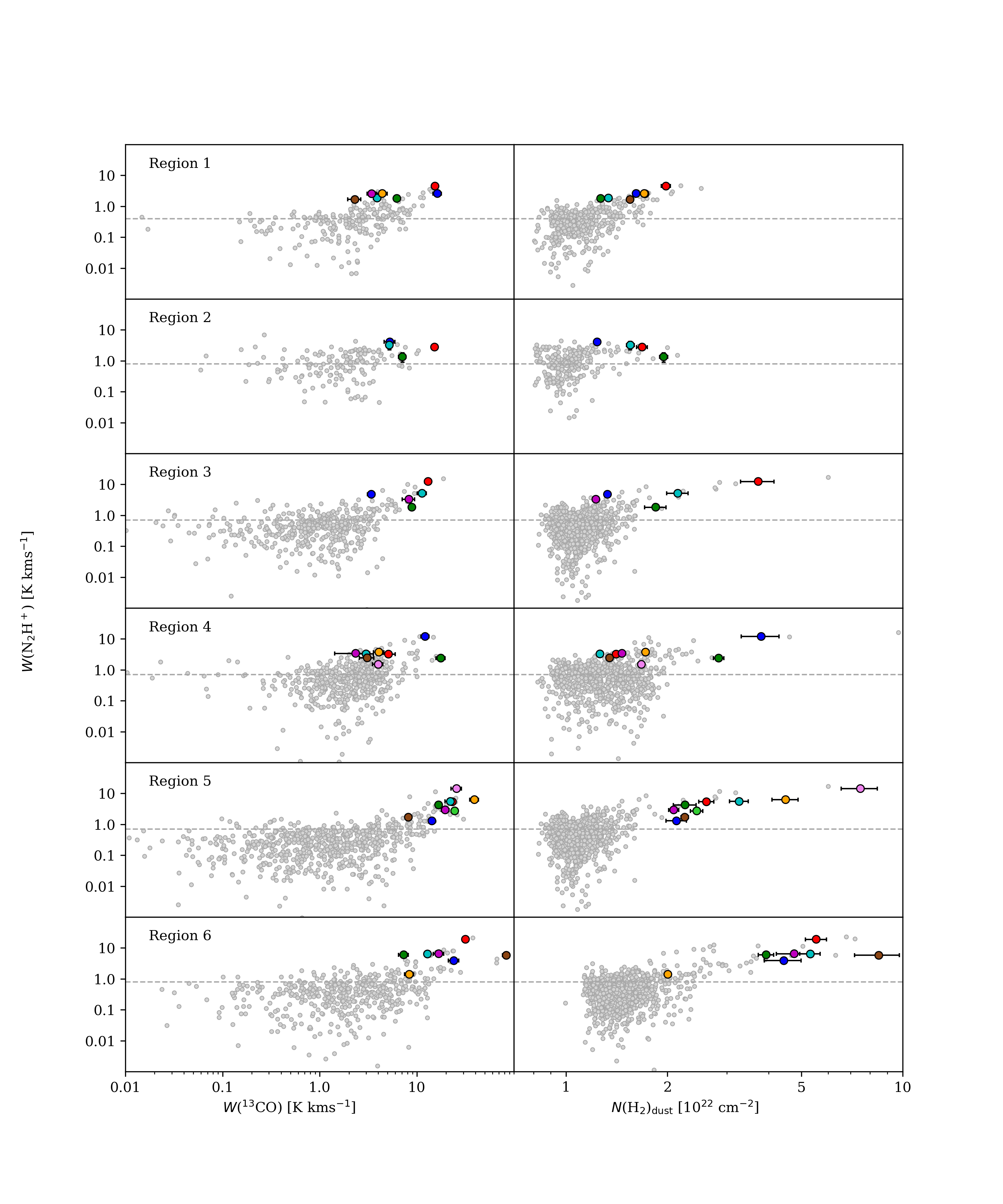}
    \caption{Comparison of the beam-averaged integrated intensities (grey circles) of $^{13}$CO(3$-$2), N$_2$H$^+$(1$-$0) and the Hi-GAL-based column density of H$_2$ on each observed region. The average values of the N$_2$H$^+$ clusters are plotted with colours corresponding to the ellipses on Fig. \ref{fig:region1_mapcl} and Fig. \ref{fig:region5_mapcl}. Dotted horizontal lines mark the noise level of the $W$(N$_2$H$^+$) maps measured as the rms noise in an emission-free area.}
    \label{fig:regions_plot}
\end{figure*}

We derived the beam-averaged values of $f'_{\rm DG}$ and $h_{\rm N_2H^+}$ in each region using each pixel where the detected N$_2$H$^+$(1$-$0) emission was above zero. Figure \ref{fig:allregion_fh}a shows the distribution of these values for every region. The average values of $f'_{\rm DG}$ are systematically higher for Region 2, 3 and 4 (0.50\,$\pm$0.07, 0.47\,$\pm$\,0.06, 0.33\,$\pm$\,0.03, respectively) than for Region\,1 and the two arm filament regions, 5 and 6 (with an average of 0.17\,$\pm$\,0.02, 0.18\,$\pm$\,0.02, 0.19\,$\pm$\,0.03, respectively). This is also the case when looking at the averages of $h_{\rm N_2H^+}$, and the difference also exists when considering the medians of both parameters.

The regions in the arm filament show higher average $W$($^{13}$CO) and $N$(H$_2$)$_{\rm dust}$ than the inter-arm filament (see Fig. \ref{fig:allregion_fh}b), while such difference is less apparent in the N$_2$H$^+$(1$-$0) emission which is similarly bright in the two filaments. A correlation between the average $W$($^{13}$CO) of a region and its average $f'_{\rm DG}$ by eye shows that the regions with fainter $^{13}$CO(3$-$2) emission have higher $f'_{\rm DG}$ and the arm filament regions with brighter $^{13}$CO emission (along with Region\,1) exhibit a lower $f'_{\rm DG}$. This is also true when looking at $f'_{\rm DG}$ and $h_{\rm N_2H^+}$ versus $N$(H$_2$)$_{\rm dust}$: the denser, CO-brighter arm filament shows a lower dense gas fraction. We quantify these correlations by calculating the Spearman rank-order correlation coefficient between the variables which assesses how well the relation between the variables can be described with a monotonous function. We find Spearman coefficients for the average $W$($^{13}$CO), $W$(N$_2$H$^+$) and $W$(NH$_2$)$_{\rm dust}$ versus average $f'_{\rm DG}$ of 0.83 (p-value: 0.04), 0.66 (0.16), and -0.54 (0.27), respectively. The same assessment for the average $W$($^{13}$CO), $W$(N$_2$H$^+$) and $N$(H$_2$)$_{\rm dust}$ versus average $h_{\rm N_2H^+}$ gives -0.60 (0.21), 0.89 (0.02), and -0.60 (0.21), respectively. Only the correlation between the average $^{13}$CO brightness and $f'_{\rm DG}$, and between $W$(N$_2$H$^+$) and $h_{\rm N_2H^+}$ seem significant, although all coefficients are above 0.5.

Figure \ref{fig:regions_plot} shows the correlations of the beam-averaged integrated intensities of N$_2$H$^+$(1$-$0) and $^{13}$CO(3$-$2), alongside the Hi-GAL based $N$(H$_2$)$_{\rm dust}$ for each region. Above the noise level of the $W$(N$_2$H$^+$) maps the integrated intensity of N$_2$H$^+$(1$-$0) generally increases with both increasing $W$($^{13}$CO) and $N$(H$_2$)$_{\rm dust}$. The Spearman coefficients between $W$($^{13}$CO) and $W$(N$_2$H$^+$) in Region 1$-$6 are 0.54, 0.22, 0.27, 0.28, 0.36, and 0.32 (with p-values\,<\,0.01), and between $N$(H$_2$)$_{\rm dust}$ and $W$(N$_2$H$^+$) are 0.44, 0.32, 0.23, 0.12, 0.26 and 0.28 (with p-values\,<\,0.01). The data of Region\,1 shows the highest coefficient.

\subsection{The parsec-scale clustering of N$_2$H$^+$(1$-$0) emission}

Fig. \ref{fig:regions_plot} also shows that the identified N$_2$H$^+$ clusters all appear above certain $W$(N$_2$H$^+$), $W$($^{13}$CO), and $N$(H$_2$)$_{\rm dust}$ thresholds. The $W$($^{13}$CO) threshold seems to be larger for Regions\,5$-$6 than for the inter-arm filament regions, around 8\,K\,kms$^{-1}$ instead of 2$-$5\,K\,kms$^{-1}$. The $N$(H$_2$)$_{\rm dust}$ threshold for the arm filament regions is also higher, 2\,$\times$\,10$^{22}$\,cm$^{-2}$, while it is around 1.2\,$\times$\,10$^{22}$\,cm$^{-2}$ for Regions\,1$-$4.

The values of $f'_{\rm DG}$ and $h_{\rm N_2H^+}$ computed for the N$_2$H$^+$ clusters give $f'_{\rm DG}$\,=\,0.46, 0.44, 0.68, and 0.80 on average for the clusters in Regions\,1$-$4 and 0.22 and 0.39 for Regions\,5$-$6, thus, an average of 0.60 for the inter-arm filament and 0.31 for the arm filament (we did not consider R1CL7 and R4CL8 in this computation). This agrees well with the results gained from the beam-averaged values before. The average $h_{\rm N_2H^+}$ values are 1.52, 1.93, 2.60 and 2.0\,K\,kms$^{-1}$/10$^{22}$\,cm$^{-2}$ for Regions\,1$-$4 (an average of 2.0\,K\,kms$^{-1}$/10$^{22}$\,cm$^{-2}$) and 1.43 and 1.40 for Regions\,5$-$6. Both the average $f'_{\rm DG}$ and the $h_{\rm N_2H^+}$ values of all beams and the averages of the N$_2$H$^+$ clusters in the two filaments suggest higher dense gas fraction in the inter-arm than in the arm.
Table \ref{tab:region_clusters} contains all the parameters (extent, average spectral parameters, integrated intensities, and column densities) for our identified N$_2$H$^+$ clusters.

We associate the clumps from the Hi-GAL catalogue by \citet{elia2021} to our N$_2$H$^+$ clusters to investigate their physical properties further. Out of our 40 N$_2$H$^+$ clusters 32 have associated objects in the Hi-GAL sample, 17 in the inter-arm filament and 15 in the arm filament. The catalogue lists the classification of the clumps based on the existence of 70\,$\micron$ Herschel-emission and establishes three groups: protostellar, pre-stellar (starless but bound), and starless (starless and unbound). In the inter-arm filament, 11 of the N$_2$H$^+$ clusters that could be associated with Hi-GAL clumps are protostellar (65\%), 5 are pre-stellar and 1 is starless (35\%). In the arm filament, all 15 associated clumps are protostellar. For comparison, the total number of Hi-GAL clumps in the N$_2$H$^+$-mapped area of the arm filament is 112, from which 21.5\% are starless/pre-stellar and 78.5\% are protostellar, and in the inter-arm filament it is 150, from which 53\% are starless/pre-stellar, 47\% are protostellar. 

\begin{figure*}
    \centering
    \includegraphics[width=.45\linewidth]{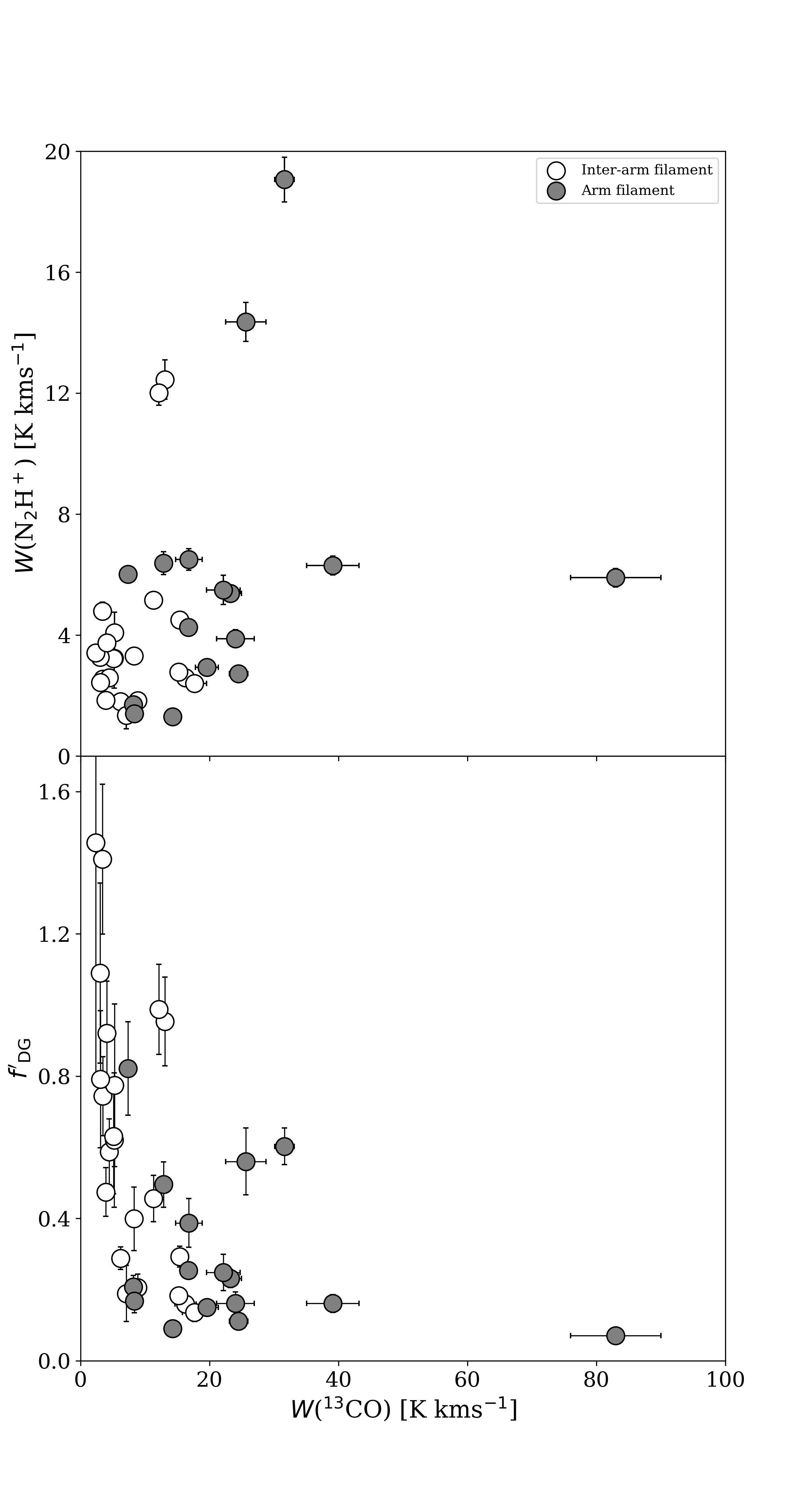}
    \includegraphics[width=.45\linewidth]{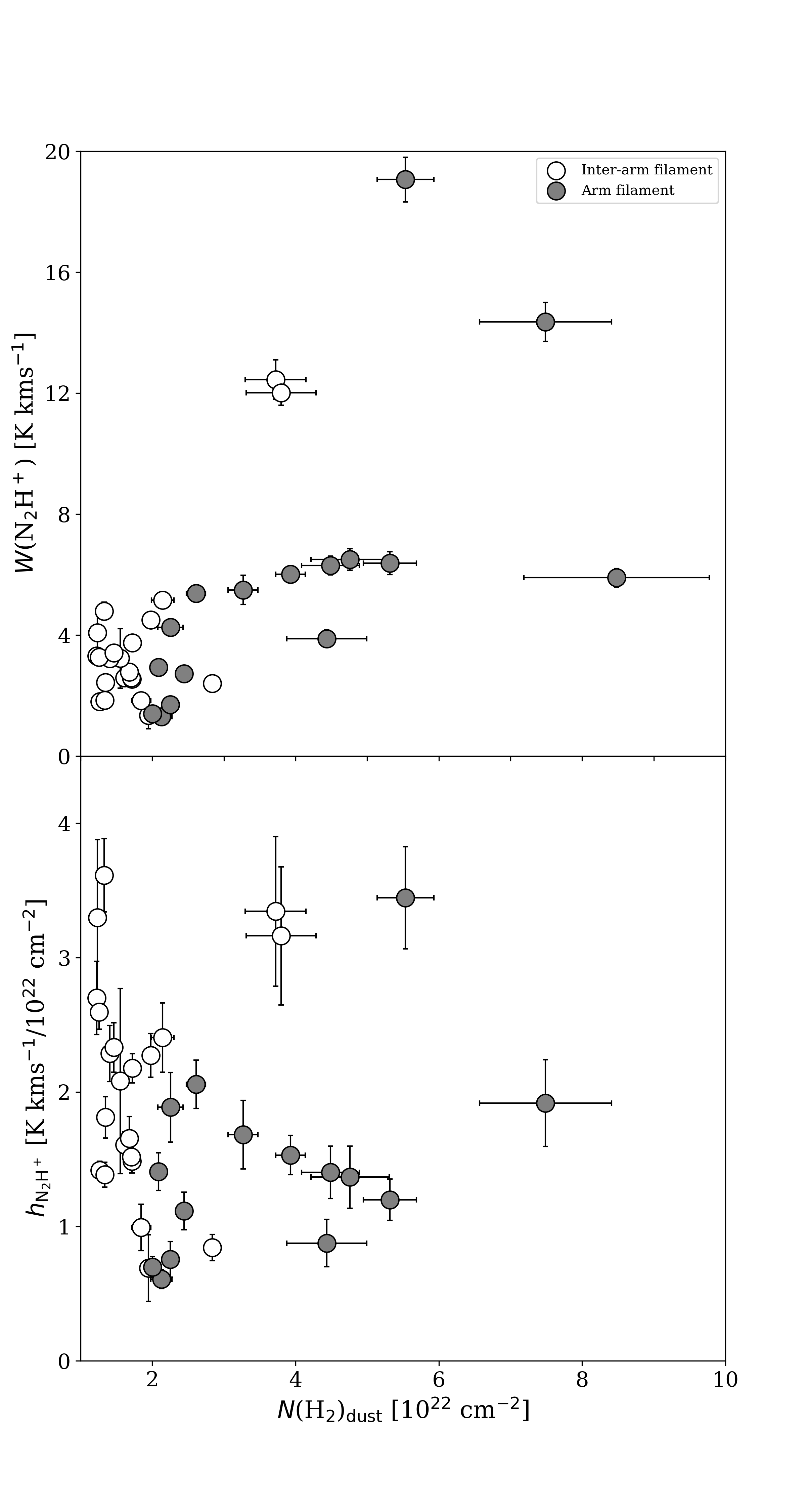}
    \caption{The correlation of the average N$_2$H$^+$(1$-$0) integrated intensity, $^{13}$CO(3$-$2) integrated intensity, H$_2$ column density, and the two types of dense gas fraction measures in the N$_2$H$^+$ clusters of the arm and inter-arm filaments.}
    \label{fig:allregion_clusters}
\end{figure*}

Figure \ref{fig:allregion_clusters} shows the correlations of the average $W$(N$_2$H$^+$) versus average $W$($^{13}$CO) and $N$(H$_2$)$_{\rm dust}$ for all the inter-arm and arm filament N$_2$H$^+$ clusters. Inter-arm clusters appear at lower $W$($^{13}$CO) values than arm clusters and their maximum $^{13}$CO brightness is also lower. Inter-arm clusters have roughly constant average $W$(N$_2$H$^+$) at low $N$(H$_2$)$_{\rm dust}$ but with increasing column densities, a couple of clusters appear at higher $W$(N$_2$H$^+$) as well. In contrast, arm filament clusters show generally higher $N$(H$_2$)$_{\rm dust}$ values and their average $W$(N$_2$H$^+$) starts rising already at 2\,$\times$\,10$^{22}$\,cm$^{-2}$. The behaviour of $h_{\rm N_2H^+}$ versus $N$(H$_2$)$_{\rm dust}$ differs from that of $f'_{\rm DG}$ versus $W$($^{13}$CO): while the latter shows a general decreasing tendency with $^{13}$CO(3$-$3) brightness, the former is mostly constant with increasing $N$(H$_2$)$_{\rm dust}$.

Arm clusters are all protostellar where association was possible, and on average, they show bright $^{13}$CO emission, high $N$(H$_2$)$_{\rm dust}$, and generally low $f'_{\rm DG}$ and $h_{\rm N_2H^+}$. In the inter-arm, there is no significant difference between the $^{13}$CO emission of protostellar or starless/prestellar clumps, however the $N$(H$_2$)$_{\rm dust}$ of starless/prestellar clumps tends to be lower than for the protostellar ones, and only protostellar clumps reach really high N$_2$H$^+$ brightness.

\section{Discussion}
\label{sec:disc}

\subsection{The observed scales in context}

\begin{figure}
    \centering
    \includegraphics[width=.85\linewidth]{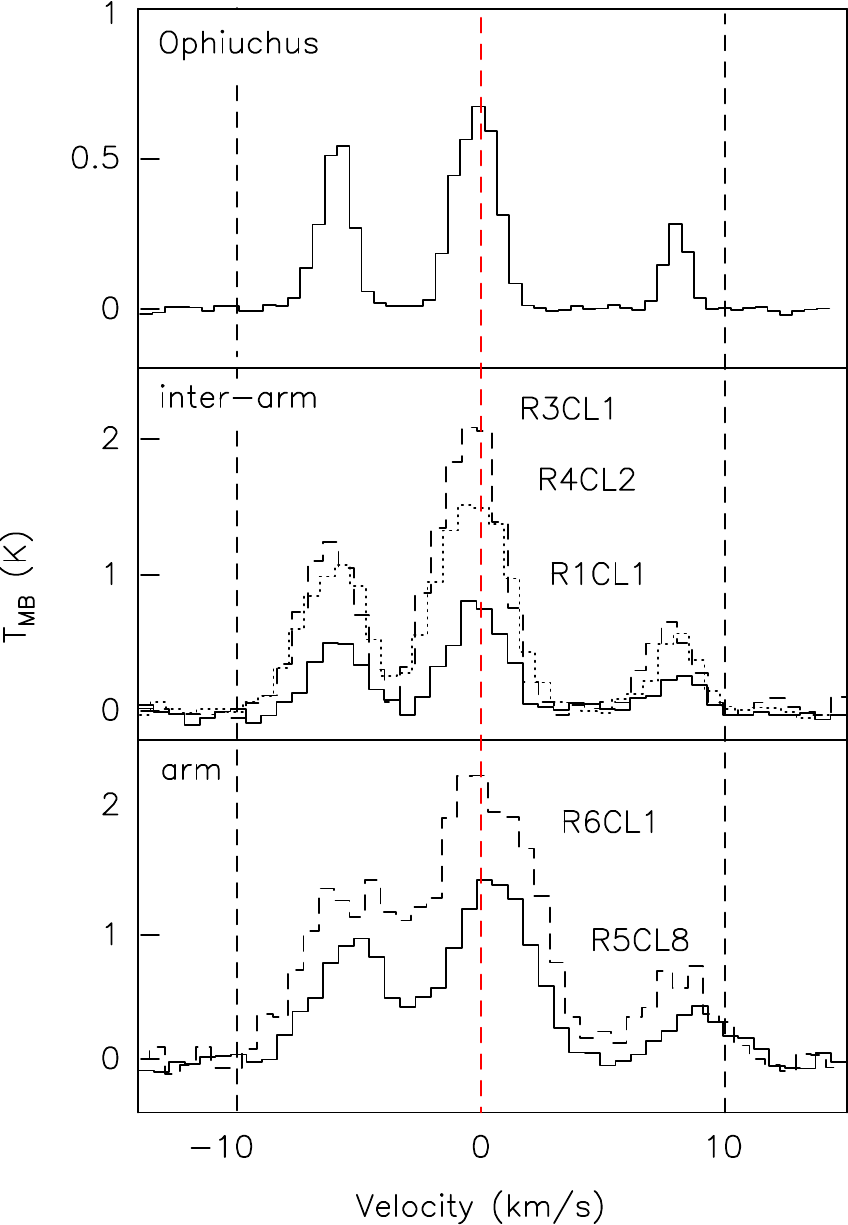}
    \caption{The average N$_2$H$^+$(1$-$0) spectrum of the Ophiuchus main cloud based on the observations of \citet{andre2007} and the average spectra of three inter-arm and two arm filament clusters. All spectra were centered to $v_{\rm LSR}$\,=\,0 for simple comparison purposes. The spectrum of Ophiuchus was also smoothed to be presented with a similar velocity resolution (0.5\,kms$^{-1}$) as the other spectra (0.63\,kms$^{-1}$).}
    \label{fig:ophi}
\end{figure}

To better interpret our results, it is advantageous to put the spatial scales of our observations in perspective. The study of Ophiuchus also using the IRAM\,30m telescope by \citet{andre2007} provides a good opportunity for comparison. They carried out a mapping of N$_2$H$^+$(1$-$0) towards the protostellar condensations in the main Ophiuchus molecular cloud to examine the kinematics of the region. Their Fig. 7 shows the large-scale view of their studied cloud containing 41 condensations. The entire region is roughly 1\,pc in diameter and thus would fit into one of our N$_2$H$^+$ clusters, for example R1CL1, R3CL1, or R4CL2 in the inter-arm filament or R5CL8 and R6CL6 in the arm filament (the brightest and largest clusters). A comparison of these spectra can be seen in Figure \ref{fig:ophi}. The N$_2$H$^+$(1$-$0) lines appear with comparable intensities and linewidths to our measurement set, with Ophiuchus showing $\Delta v_{\rm HFS}$\,=\,1.0\,kms$^{-1}$, the shown inter-arm clusters 1.8$-$2.5\,kms$^{-1}$ and the shown arm clusters 3.0$-$3.7\,kms$^{-1}$. While the variation in velocity dispersion between the different regions may originate from both intrinsic structure and from integrating many different cloud structures into the beam, the similarity suggests that there is a good chance our N$_2$H$^+$ clusters are star forming clouds similar to Ophiuchus, at a larger distance and all their line emission integrated into 1-2 beams.

\subsection{Line emission and the dense gas fraction in large-scale galactic structures}

Different samples of GMFs similar to ours but identified by various other methods such as by \citet{ragan2014}, \citet{abreu2016}, \citet{zucker2015}, \citet{wang2015}, and \citet{wang2016} were found to vary in their physical parameters (aspect ratio, linear mass and column density) signifying true differences in their physical states, and potentially origin and role in the galactic structure \citet{zucker2018}. The filaments studied by us have full lengths at the middle of the range they investigated (11$-$269\,pc) and are more similar to their so-called elongated giant molecular cloud specimens. 

The dense gas mass fraction for these samples of filaments were calculated in various ways. For the GMFs in the sample of \citet{ragan2014}, the dense gas mass fraction was determined from the ratio of dense gas mass measured by the ATLASGAL 870$\micron$ maps divided by the total gas mass measured by the GRS $^{13}$CO survey. They found a dense gas mass fraction of 1.6$-$12\% for their inter-arm objects which is consistent with values for other large molecular filaments, local star-forming clouds, high-mass regions and local star-forming clumps \citep{battersby2014, kainulainen2009, kainulainen2013, lada2010, lada2012, johnston2009, battisti2014}. They also found that the dense gas mass fraction decreases with increasing galactocentric radius, increases towards the mid-plane, and the GMF closest to a spiral arm shows the highest value. The work by \citet{abreu2016} calculates the dense gas mass fraction the same way and finds values between 1.5$-$14\% with one filament related to a massive HII complex showing 37\%. They see no correlation with distance from the midplane. In their sample, arm filaments also show higher dense gas mass fraction, although most of their objects are in the arm. In contrast with this, we find higher values of $f'_{\rm DG}$ for our inter-arm filament than for the Sagittarius-arm filament, both when looking at general trends and in the extracted N$_2$H$^+$ clusters (with the caveat that our method of deriving dense gas fraction is different than in the studies mentioned above). We note that the first GMF identified, "Nessie", a structure inside a spiral arm, shows around 50\% dense gas mass fraction, derived from HNC\,(1-0) emission tracing the dense gas \citep{jackson2006, goodman2014}.

The brightness of $^{13}$CO(3$-$2) and dust emission shows differences between our two filaments with the arm filament generally showing higher values, while the N$_2$H$^+$(1$-$0) brightness is generally similar in the two environments. \citet{moore2012} examined cloud masses and YSO content in the GRS survey region and concluded that most of the observed increase in star-formation rate density associated with spiral arms is due to source crowding and there is little or no increase in star-formation efficiency (SFE) on kiloparsec scales. \citet{eden2013} studied the GRS clouds by associating dense clumps to them from the BGPS \citep[Bolocam Galactic Plane Survey, ][]{aguirre2011} sample and found no evidence of difference in clump-formation efficiency (CFE, which can be thought of as a precursor-analogue of SFE) between inter-arm and spiral-arm regions. \citet{schinnerer2017} performed a multi-wavelength study of a spiral arm segment in M51 and found no variation in GMC parameters between arm and inter-arm clouds, and no evidence of a coherent SF onset mechanism that can be solely associated with the arms. In contrast, in our sample, both $^{13}$CO and dust IR emission average and maximum brightness show differences between inter-arm and arm regions. It looks like there is more dense gas to form stars in the inter-arm than in the spiral-arm cloud but since the N$_2$H$^+$(1$-$0) emission does not show significant differences in strength between the two environments, the derived dense gas fraction values and their tendencies are controlled by our method of measuring the total gas, thus the $^{13}$CO integrated intensities and the H$_2$ column densities. Measuring total gas with $^{13}$CO, the derived $f'_{\rm DG}$ values show large differences in the two filaments and also a decreasing tendency versus $^{13}$CO brightness (see Figure \ref{fig:allregion_clusters}). These differences and tendencies diminish when deriving total gas using the IR continuum, thus H$_2$ column densities. We also find only protostellar Hi-GAL clumps on our mapped area in the arm filament, while the ratio of protostellar objects in the inter-arm is much lower.

The Hi-GAL-based dust temperatures on the maps derived by PPMAP do not seem to be very different for the two filaments, however some difference can be seen in the average temperatures of the Hi-GAL clumps computed by \citet{elia2021} where the clumps in the arm are warmer. The average $^{13}$CO(3$-$2) brightness inside arm filament sources also shows a positive correlation with the temperatures of the clumps, while the N$_2$H$^+$(1$-$0) brightness does not. The $^{13}$CO(3$-$2) brightness$-T_{\rm dust}$ correlation does not apply in the inter-arm filament.

\subsection{How do we measure the dense gas fraction?}

\citet{pety2017} concludes in their investigation of Orion\,B that despite the similar critical densities of N$_2$H$^+$ and HCO$^+$, their emitting behaviour is different. In their dataset, N$_2$H$^+$ is only emitting from dense regions while HCO$^+$ is not. N$_2$H$^+$ only survives in regions where the electron abundance is low (which prevents dissociative recombination) and where CO is frozen onto dust grains (which prevents proton transfer to CO). Thus, they propose to use N$_2$H$^+$/$^{12}$CO as a dense gas fraction measure. They measure the flux ratios of species e.g. $^{12}$CO/N$_2$H$^+$ which is 900 in Orion\,B. Collected values from the literature are: 40 in ULIRGs and 59$-$167 to LMC \citep{nishimura2016}, 225 for M51\,P2 \citep{watanabe2014}, 41$-$68 for AGNs and 100$-$325 for starburst galaxies \citep{aladro2015}. The value of $^{13}$CO/N$_2$H$^+$ in our N$_2$H$^+$ clusters is generally around 1$-$5 which should result in a $^{12}$CO/N$_2$H$^+$ of 50$-$250 working with a $^{12}$CO\,=\,45$-$55\,$\times$\,$^{13}$CO from $^{12}$CO/$^{13}$CO\,=\,5.41R$_{\rm gal}$+19.3 at R$_{\rm gal}\,\approx\,$5$-$7 by \citet{milam2005}. However, since $^{12}$CO is usually optically thick, affecting the measured line intensity, towards real star forming sites the $^{12}$CO/$^{13}$CO ratio is closer to 10 \citep{tafalla2023}. The conversion is also dependent on the specific transitions used. $^{13}$CO(3$-$2)/$^{13}$CO(1-0) is probably lower than 1, increasing $^{12}$CO/N$_2$H$^+$, thus making this comparison riddled with uncertainties. \citet{pety2017} measures a filling factor of 2.4\% and an abundance of 2\,$\times$\,10$^{-11}-$3\,$\times$\,10$^{-9}$ for N$_2$H$^+$ in Orion\,B versus our 8.6\% and 1.5\,$\times$\,10$^{-10}-$1.06\,$\times\,$10$^{-9}$.

\citet{kauffmann2017} investigated Orion\,A for the LEGO project using several molecular species (e.g. CO isotopologues, CN, CCH, HCN, N$_2$H$^+$). They measured the behaviour of $h_{\rm Q}$\,=\,$W$(Q)/$A_{\rm V}$ versus $A_{\rm V}$ for the molecules and saw a trend for CO isotopologues, C$_2$H, HCN, and CN where $h_{\rm Q}$ is high at low $A_{\rm V}$, the value increases to a maximum around 5$-$20 mag of $A_{\rm V}$ then steadily decreases. In contrast, the $h_{\rm Q}$ values for N$_2$H$^+$ begin at nearly zero at low $A_{\rm V}$ then start rising around 10\,mag to reach a maximum or a plateau around 100\,mag. We measure $A_{\rm V}$ around 10$-$16 in our current study and see a roughly constant behaviour of $h_{\rm N_2H^+}$ versus $N$(H$_2$)$_{\rm dust}$ (with scatter) as predicted by \citet{priestley2023a}.

\citet{barnes2020} presented a map of the massive star forming region W49 (d\,$\approx$\,11\,kpc) with the IRAM 30\,m telescope in the emission of several molecular species for the LEGO project. They also use the $h_Q$ measure after \citet{kauffmann2017}. Their Figure 7 shows an increase of $W$(N$_2$H$^+$) versus $N$(H$_2$)$_{\rm dust}$ starting slightly below 10$^{22}$\,cm$^{-2}$ from a plateau that appears on our figures for our targeted filament regions as well (Figure \ref{fig:regions_plot}). In their sample the CS, CH$_3$OH, HCO$^+$ and N$_2$H$^+$ integrated intensities rise the quickest with increasing $N$(H$_2$)$_{\rm dust}$. Their values of $h_{\rm Q}$ show a quick rise with $N$(H$_2$)$_{\rm dust}$ and a turnoff point around 5\,$\times$\,10$^{22}$\,cm$^{-2}$ (similarly to e.g. HCO$^+$, CN, CS, HCN, HNC) and they found that their results regarding $h_{\rm Q}$ are in good agreement with \citet{pety2017}. This was a surprising conclusion because of the very different resolution used (2$-$3\,pc versus their 0.05\,pc). 

Comparing the results in W49 by \citet{barnes2020} with those of Orion\,B by \citet{pety2017} further comes up with interesting differences: for Orion\,B, $h_{\rm Q}$ (the emission efficiency as they call it) for HCN peaks at a lower column density than for N$_2$H$^+$, while in W49 they found the two peak points comparable, with N$_2$H$^+$ even potentially peaking at lower column densities. To investigate this, \citet{barnes2020} assessed how $h_{\rm Q}$ varies as a function of dust temperature and column density simultaneously. They found that the majority of the line emission of HCN is towards moderate-high temperatures (around 30$-$35\,K), while for N$_2$H$^+$ it is at around the lowest observed temperatures at the highest column density within the observed parameter space (around 20\,K). There is also another peak for N$_2$H$^+$ at higher temperatures. They explain the difference with the different chemical formation pathways of the two species which can result in them favouring different temperature regimes and can have important effects on the gas that we may trace using them.

As mentioned in Section \ref{sec:dgmf}, the different ways of computing dense gas fraction for our structures will introduce varying biases regarding the behaviour of tracers in the interstellar radiation field (shielding, depletion, photo-dissociation, etc.) but even so, the detection of higher dense gas fraction in the inter-arm regions in our data seems counter-intuitive. One explanation may be the higher protostellar clump fraction in the arm regions. Newly-forming stars will heat the surrounding gas and dust, which can enhance the CO emission by both desorbing frozen-out molecules and so enhancing the abundance, and due to the higher gas temperatures, enhancing the excitation temperature of the molecule \citep{clark2012}. We would therefore expect the $^{13}$CO brightness to correlate with the Hi-GAL temperature, which does occur in the arm filament regions, as mentioned. N$_2$H$^+$ is destroyed efficiently by gas-phase CO, so the enhanced CO abundance may to some extent counteract the increased excitation for this molecule, which is consistent with the observed lack of correlation between N$_2$H$^+$ brightness and temperature. \citet{tafalla2023} recently found that the amount of CO emission per unit column density is higher in Orion A than in the Perseus and California clouds, while the N$_2$H$^+$ emission is unchanged. They attribute this to the generally higher temperatures in the actively-star-forming Orion complex, which is consistent with the scenario outlined above.

The inter-arm clouds could also suffer from the intense photo-dissociation of CO since the ISRF "seen" by the gas is G$_0$\,$\times$\,e$^{-A_{\rm V}}$ and the exponential term is potentially more important for the gas on the outskirts of the cloud where most of the mass is located. Inside the dense regions, the term should be a small number, thus irrelevant, explaining the missing CO. The G$_0$ term can also be much higher in the inter-arm than in the arm, since the high-mass star forming regions along the edge of the nearby arm(s) might shine into the inter-arm region in a less interrupted way than in the clouds inside the arm where the arm material can shield inner areas \citep{porter2017, clark2019}. The N$_2$H$^+$ brightness can also be boosted by embedded star formation. The PPMAP dust temperature maps show both the inter-arm and arm filament around 20\,K, however the separately derived $T_{\rm dust}$ values for Hi-GAL clumps of \citet{elia2021} based on their own fitting procedure show the average $T_{\rm dust}$ of Hi-GAL/N$_2$H$^+$ clusters in the inter-arm with 15.3\,K while in the arm, 20.4\,K. However, we must consider the many uncertainties in deriving the large-scale dust temperature maps used in these results e.g. foreground, background, spectral index, the usage of single-temperature fits, and the probability distribution in the results of the PPMAP procedure.

Another way to explain higher $^{13}$CO brightness in the arm filament is the presence of CO-bright material that is not associated with star formation, but rather forms in transient, dense shocks. These will likely get sheared apart by the very motions that create them, while in the inter-arm, this would happen less frequently or not at all: most of the CO that wee see is likely to be associated with bound clouds since they formed in the arm and managed to survive.

We also call attention to the assumptions used during our calculation of the optical depth, excitation temperature and column density of N$_2$H$^+$ that might introduce errors in the derived parameters e.g. optically thin conditions were assumed but as the calculations proved, the N$_2$H$^+$(1$-$0) emission is optically thick, and a filling factor of 1 was assumed, however this might be an overestimation. Additionally, it is worth mentioning here again the results by \citet{barnes2020} who find that the emission efficiency of N$_2$H$^+$ in W49 peaks at two locations in the $T_{\rm dust}-N$(H$_2$)$_{\rm dust}$ space. The stronger peak is at lower temperatures, while the emission is less efficient but significant in warmer environments. This might also influence the amount of N$_2$H$^+$ we detect in our regions in different environmental characteristics.

\subsection{Extragalactic outlook}

Regarding the extragalactic implications of the current study, there are only a small number of extragalactic maps analysing N$_2$H$^+$(1$-$0) emission as of date, however, with the large surveys of ALMA and the IRAM telescopes (e.g. EMPIRE, ALCHEMI) near-future influx of new results is expected. \citet{harada2019} studied the circumnuclear ring of M83 with ALMA and observed different N$_2$H$^+$(1$-$0)/$^{13}$CO(1$-$0) ratios at different positions, generally between 0.1$-$0.24. They explain the variation with potentially increasing N$_2$H$^+$ abundance in a collapsing prestellar core (shocks compress gas and increase the dense gas fraction). Comparison with our results is difficult, however, due to the different observed $^{13}$CO isotopologue and the vastly different scales involved (their "clumps" are 50$-$100\,pc large, thus, around a hundred times larger than ours). Their N$_2$H$^+$ column densities are in line with the higher values we measure (a few times 10$^{12}-$10$^{13}$\,cm$^{-2}$) and their excitation temperatures are around 6\,K, which we only measure in Region\,6. \citet{eibensteiner2022} observed 2 and 3\,mm molecular lines towards the centre of NGC\,6946 with NOEMA, however, they do not analyse N$_2$H$^+$ separately, only against the emission of HCN.

\citet{jimenez2023} observed N$_2$H$^+$(1$-$0) in the spiral galaxy NGC\,6946 across different environments and found a strong correlation of its emission with HCN at kiloparsec scales. They made pointed observations of HCN and N$_2$H$^+$ and observe that although the HCN/CO ratio does decrease with galactocentric radius, N$_2$H$^+$ seems to be constant. Thus, their primary result is that there is a strong correlation of HCN and N$_2$H$^+$ through different scales and orders of magnitude. They find N$_2$H$^+$/CO values between 0.002$-$0.0047 with no significant difference between arm and inter-arm regions. Recently, the NOEMA mapping of M51 was published by \citet{stuber2024} also presenting the line ratios of HCN, N$_2$H$^+$ and $^{12}$CO, this time on smaller spatial scales. Their fit to the N$_2$H$^+$ versus HCN line emissions in the central region of the galaxy close to its AGN is in better agreement with the values computed by \citet{jimenez2023}, but differ from those when fitting values observed on pixels in the N$_2$H$^+$-bright south-western spiral arm of M51. The investigation of the N$_2$H$^+$-to-HCN ratio in our filaments will be the topic of a future paper.


\section{Conclusions} 
\label{sec:conc}

We performed large-scale on-the-fly mapping of dense gas tracers along two giant molecular filaments in the Galaxy, one associated with the Sagittarius arm and one with an inter-arm environment. We used the N$_2$H$^+$(1$-$0) transition to trace the dense gas and compare its distribution to tracers of more extended and diffuse gas, namely $^{13}$CO(3$-$2) and dust emission. The ratio of dense to diffuse gas over a fixed area, what we call the dense gas fraction and measure with two parameters, $f'_{\rm DG}$ and $h_{\rm N_2H^+}$, may be a key indicator of star formation efficiency, thus investigating their environmental differences is of utmost importance. 

Our main findings are:
\begin{enumerate}
    \item The N$_2$H$^+$(1$-$0) emission generally traces the brightest $^{13}$CO(3$-$2)-emitting and dusty features of the filaments with a filling factor of around 8.5\%. This dense gas appears in small, roughly parsec-sized clusters with no substructure on the observed scales.
    \item Evaluating the correlation of the $^{13}$CO(3$-$2) and N$_2$H$^+$(1$-$0) emission and their ratio, $f'_{\rm DG}$, an analogous expression to the dense gas mass fraction, inter-arm cloud regions show a larger fraction of dense gas on average than the Sagittarius-arm filament. This is also observable by calculating the average $f'_{\rm DG}$ of the individual N$_2$H$^+$ clusters extracted with dendrogram clustering, resulting in values of 0.60 for the inter-arm filament and 0.31 for the arm filament.
    \item Another dense gas mass fraction-analogous value, $h_{\rm N_2H^+}$, is observed to be less changeable but still showing variations between the two filaments based on the cluster-averaged values, with 2.0\,K\,kms$^{-1}$/10$^{22}$\,cm$^{-2}$ for the inter-arm and 1.41\,K\,kms$^{-1}$/10$^{22}$\,cm$^{-2}$ for the arm filament.
    \item The arm filament is generally seen brighter in $^{13}$CO(3$-$2) and dust IR emission than the inter-arm cloud. There is, however, no such significant difference in their N$_2$H$^+$(1$-$0) emission. There also exists a $N$(H$_2$)$_{\rm dust}$ threshold above which N$_2$H$^+$(1$-$0) emission starts appearing in the mapped areas. This threshold value is observed to be higher for the arm filament.
    \item N$_2$H$^+$ clusters in the arm filament are seen as more evolved since all of them are associated with clumps in the protostellar phase, while inter-arm clusters appear as both protostellar and pre-stellar/starless. We do not see a significant difference between the average dust temperatures in the arm and inter-arm on large scales, but the arm N$_2$H$^+$ clusters seem somewhat warmer.
    \item Scenarios involving enhanced CO abundance in higher temperature gas, the observed molecular clouds suffering from the photo-dissociation by the ISRF at differing levels depending on their galactic location, the appearance of less and more unbound, transient CO-clouds, or peculiarities in the emission efficiency of N$_2$H$^+$ can explain the observed discrepancy of dense gas fraction between the arm and the inter-arm. 
\end{enumerate}

\section*{Acknowledgements}

OF, SER, FDP and PCC acknowledge the support of a consolidated grant (ST/K00926/1) from the UK Science and Technology Facilities Council (STFC). This work is based on observations carried out under project number 033-17 and E02-22 with the IRAM 30m telescope. IRAM is supported by INSU/CNRS (France), MPG (Germany) and IGN (Spain). The James Clerk Maxwell Telescope has historically been operated by the Joint Astronomy Centre on behalf of the Science and Technology Facilities Council of the United Kingdom, the National Research Council of Canada and the Netherlands Organisation for Scientific Research. This research made use of Astropy \citep[][https:// astropy.org]{astropy2022}, a community-developed core Python package for Astronomy as well as the Python packages NumPy \citep[][https://numpy.org]{oliphant2006}, SciPy \citep[][https://scipy.org]{scipy2020}, Matplotlib \citep[][https://matplotlib.org]{hunter2007} and astrodendro (http://www.dendrograms.org). This publication has also made use SAOImageDS9 \citep[][http://ds9.si.edu]{joye2003}, an astronomical imaging and data-visualization application and the CLASS and GREG software packages of GILDAS (https://www.iram.fr/IRAMFR/GILDAS).

\section*{Data Availability}

The data underlying this article will be shared on reasonable request to the corresponding author.



\bibliographystyle{mnras}
\bibliography{feher_n2h_bib} 




\appendix

\section{Deriving physical parameters from the N$_2$H$^+$ spectra}
\label{app:n2h}

Following the method used by \citet{purcell2009} we derived the full optical depth of the transition in a pixel-by-pixel manner using the HF components of N$_2$H$^+$(1$-$0) line. We perform the calculations only on pixels where the S/N of both group\,2 and group\,1 was higher than 3. We note here that as described before, the weakest group\,1 of the N$_2$H$^+$(1$-$0) HFS was often not detected which influences our ability to calculate $\tau_{\rm N_2H^+}$. Thus, occasionally, even when the main group of the transition was detected with a high S/N, we do not have a good estimation of the optical depth and the N$_2$H$^+$ column density (e.g. towards the fairly bright cluster R6CL7). 

Assuming optically thin conditions and that the linewidths and the excitation temperatures of the individual HF components are equal, the integrated intensities of the three groups that the seven HF components blend into should be in the ratio of 1:5:2. The optical depth is calculated from the integrated intensities of any two groups using:

\begin{equation}
\frac{\int T_{\rm MB,1}{\rm d}v}{\int T_{\rm MB,2}{\rm d}v}=\frac{1-{\rm e}^{-\tau_1}}{1-{\rm e}^{a\tau_1}},
\label{eq:tau}
\end{equation}

where $a$ is the expected ratio of the optical depths of the two groups, $\tau_1$ and $\tau_2$. Since \citet{caselli1995} reports anomalous excitation of HF components in group 3 (at the lowest frequencies), we determine the optical depth from the intensity ratio of group 1 and 2. 

The general equation to calculate molecular column densities from \citet{mangum2015} is:

\begin{equation}
N{\rm (total)}=\frac{3h}{8\pi^3|\mu_{\rm lu}|^2}\frac{Q_{\rm rot}}{g_{\rm u}}{\rm exp}\left(\frac{E_{\rm u}}{kT_{\rm ex}}\right) \times \left[{\rm exp} \left( \frac{h\nu}{kT_{\rm ex}}-1 \right) \right]^{-1}\int \tau_{\nu}dv,
\end{equation}

where $Q_{\rm rot}$ is the rotational partition function, $g_u$ is the degeneracy of the energy level $u$, $\mu_{lu}$ is the dipole matrix element, $E_u$ is the energy of level $u$, $\nu$ is the frequency of the transition, $h$ is the Planck constant, and $k$ is the Boltzmann constant. In the case of N$_2$H$^+$(1$-$0), assuming optically thin emission and an approximation to the integrated optical depth expression, this leads to:

\begin{equation}
\begin{split}
N({\rm N_2H^+})=1.20 \times 10^{12} (T_{\rm ex} +0.75)\frac{{\rm exp}\left(\frac{4.47}{T_{\rm ex}} \right)}{{\rm exp} \left( \frac{4.47}{T_{\rm ex}} \right)-1} \\
\times \frac{\tau}{1-{\rm exp}(-\tau)} \left[ \frac{\int T_{\rm MB} dv ({\rm km/s})}{f(J_{\nu}(T_{\rm ex})-J_{\nu}(T_{\rm bg}))} \right] {\rm cm^{-2}},
\end{split}
\label{eq:ntot}
\end{equation}

where $f$ is the beam filling fraction, $T_{\rm bg}$ is the background temperature 2.7\,K, ($T_{\rm ex}$\,+\,0.75) is an approximation of the partition function and 

\begin{equation}
    J_{\nu}(T)\,=\frac{\,\left( \frac{h\nu}{k}\right)}{\left( {\rm exp} \left( \frac{h\nu}{kT} \right)-1\right)}.
\end{equation}

In general, $T_{\rm ex}$ varies between the background radiation temperature at low densities and $T_{\rm kin}$ gas kinetic temperature at high density. A transition is said to be thermalized if $T_{\rm ex}$\,=\,$T_{\rm kin}$. In the general case of non-LTE in the ISM, most transitions of dense gas tracers are assumed to be subthermally populated. If the line is known to be very optically thick, the radiative transfer equation can be used to calculate excitation temperature analytically (because the (1-$e^{-\tau}$) term will approach 1). In the optically thin case, the observed line temperature is degenerate with $T_{\rm ex}$, $\tau$, and the filling fraction and we need to use multiple transitions using a rotation diagram. In the case of our N$_2$H$^+$(1$-$0) data (dense gas, assuming LTE, probably not very optically thick medium) we use the results obtained in CLASS by fitting the HFS.

The derived excitation temperatures are lowest in Region\,2 with 3.1\,K and highest in Region\,3 with 4.3\,K. The $\tau_{N_2H^+}$ optical depth is measured overwhelmingly above 1 (i.e. optically thick radiation) with the highest values in Region\,2 with 10 and the lowest in Region\,3 and Region\,5 with 3.1 and 3.9. Finally, N$_2$H$^+$ column densities are generally between 0.2$-$0.5\,$\times$10$^{14}$\,cm$^{-2}$ except in Region\,2 where it is higher with 1.2\,$\times$10$^{14}$\,cm$^{-2}$. The average $T_{\rm ex}$, $\tau$ and $N$(N$_2$H$^+$) for the N$_2$H$^+$ clusters can be found in Table \ref{tab:region_clusters} wherever the observed spectra was detected with a good enough S/N to compute those.

\section{Figures and Tables}

\begin{figure*}
    \centering
    \subfigure{\includegraphics[width=0.44\linewidth]{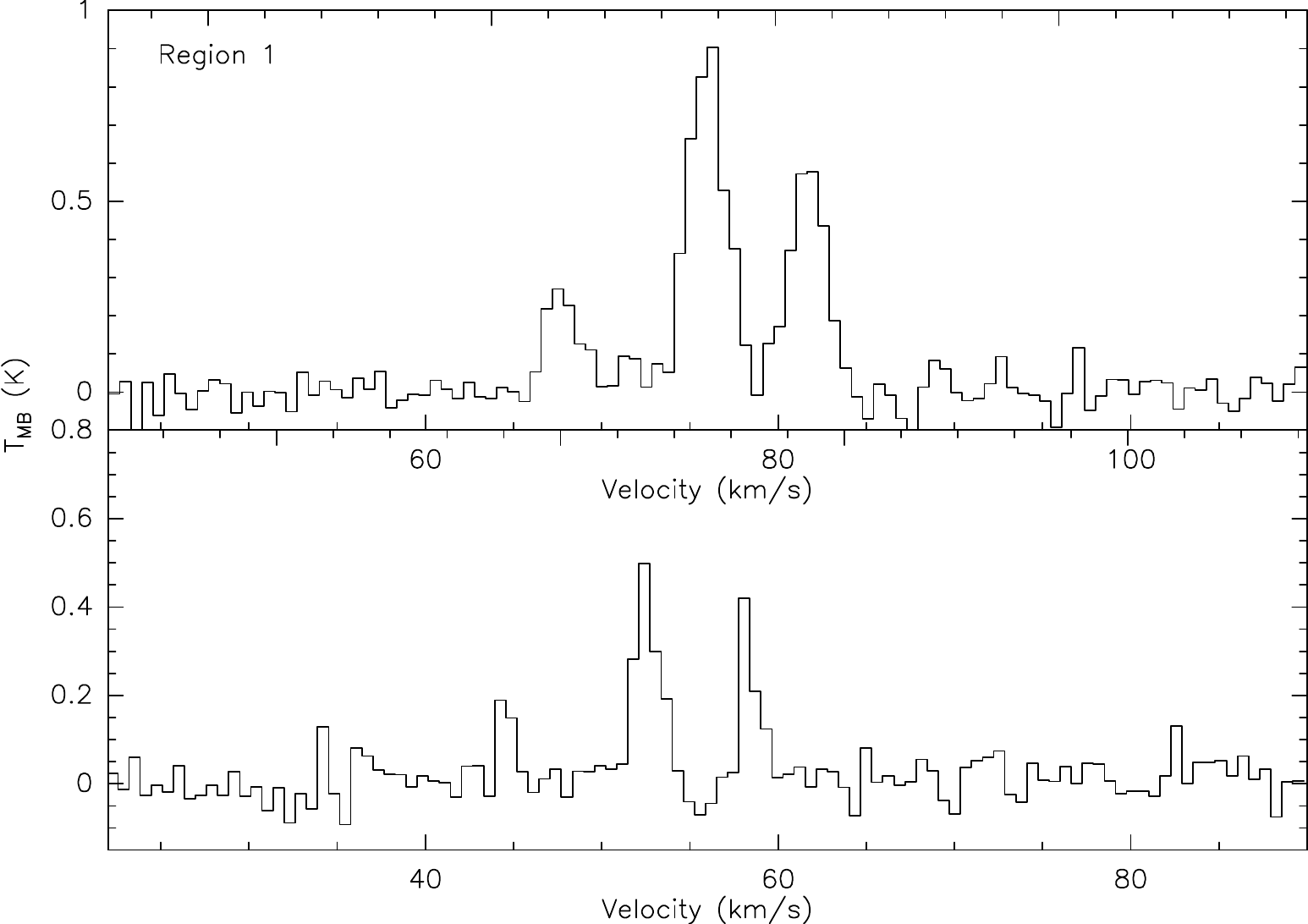}}
    \subfigure{\includegraphics[width=0.44\linewidth]{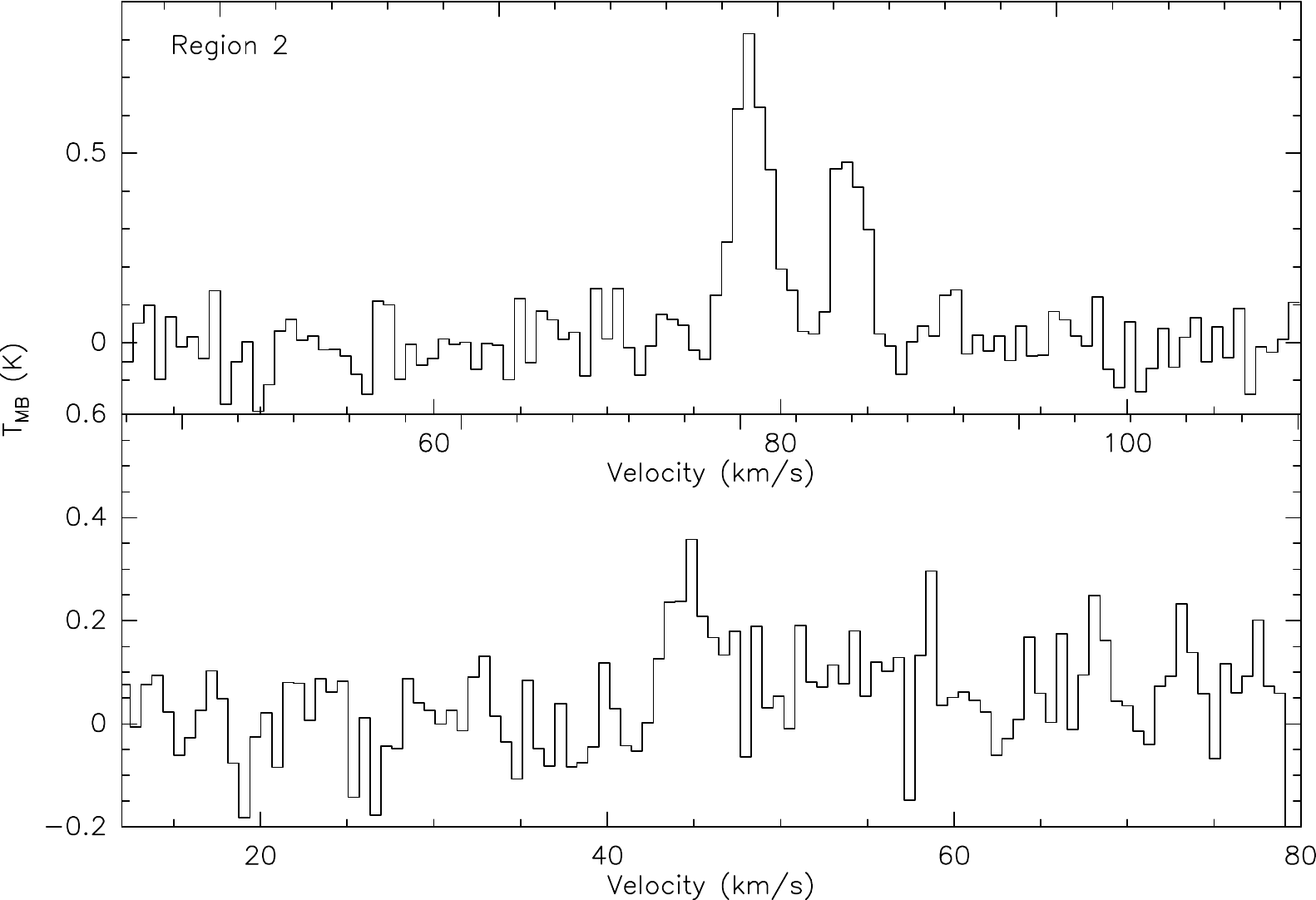}}
    \subfigure{\includegraphics[width=0.45\linewidth]{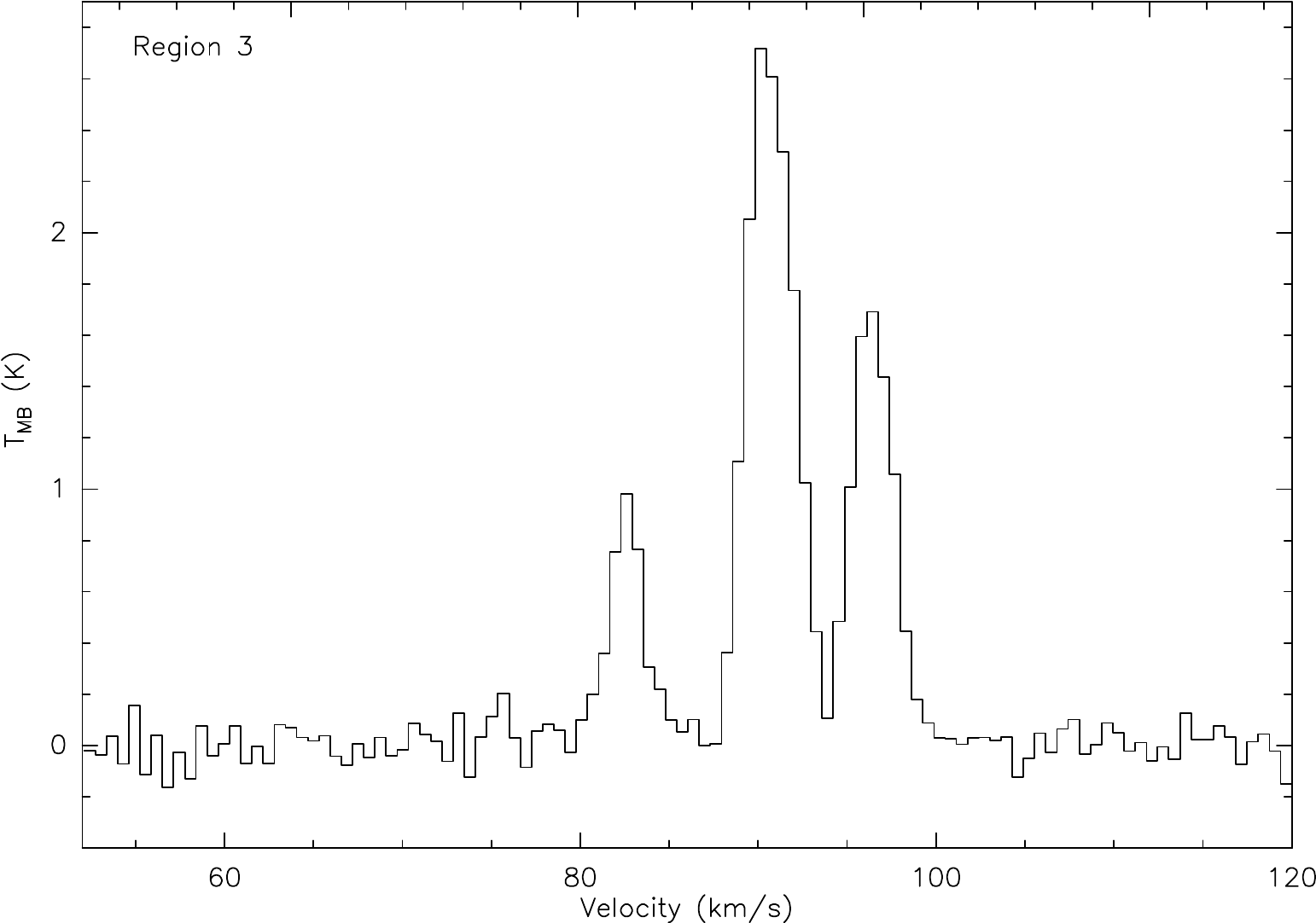}}
    \subfigure{\includegraphics[width=0.45\linewidth]{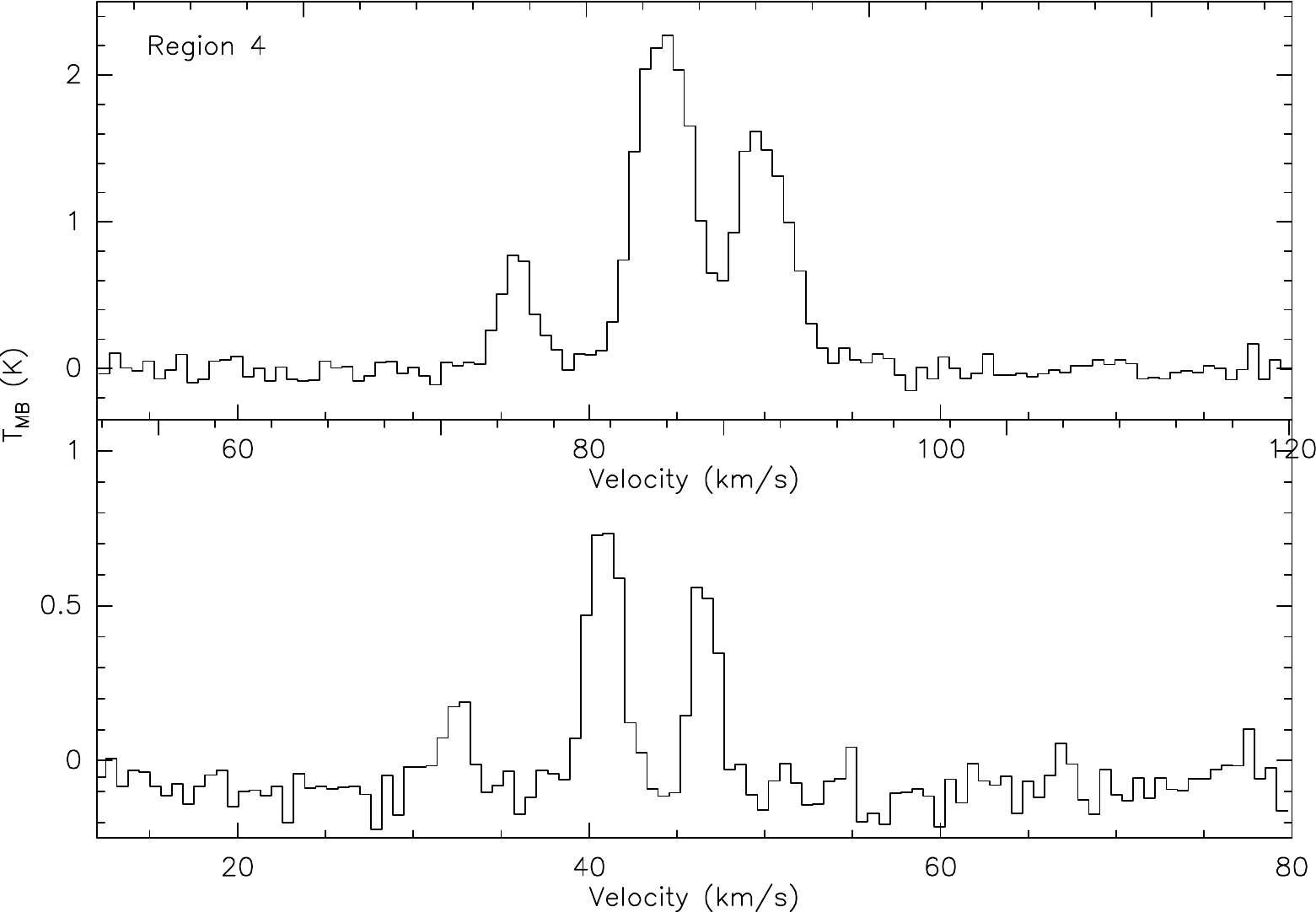}}
    \subfigure{\includegraphics[width=0.45\linewidth]{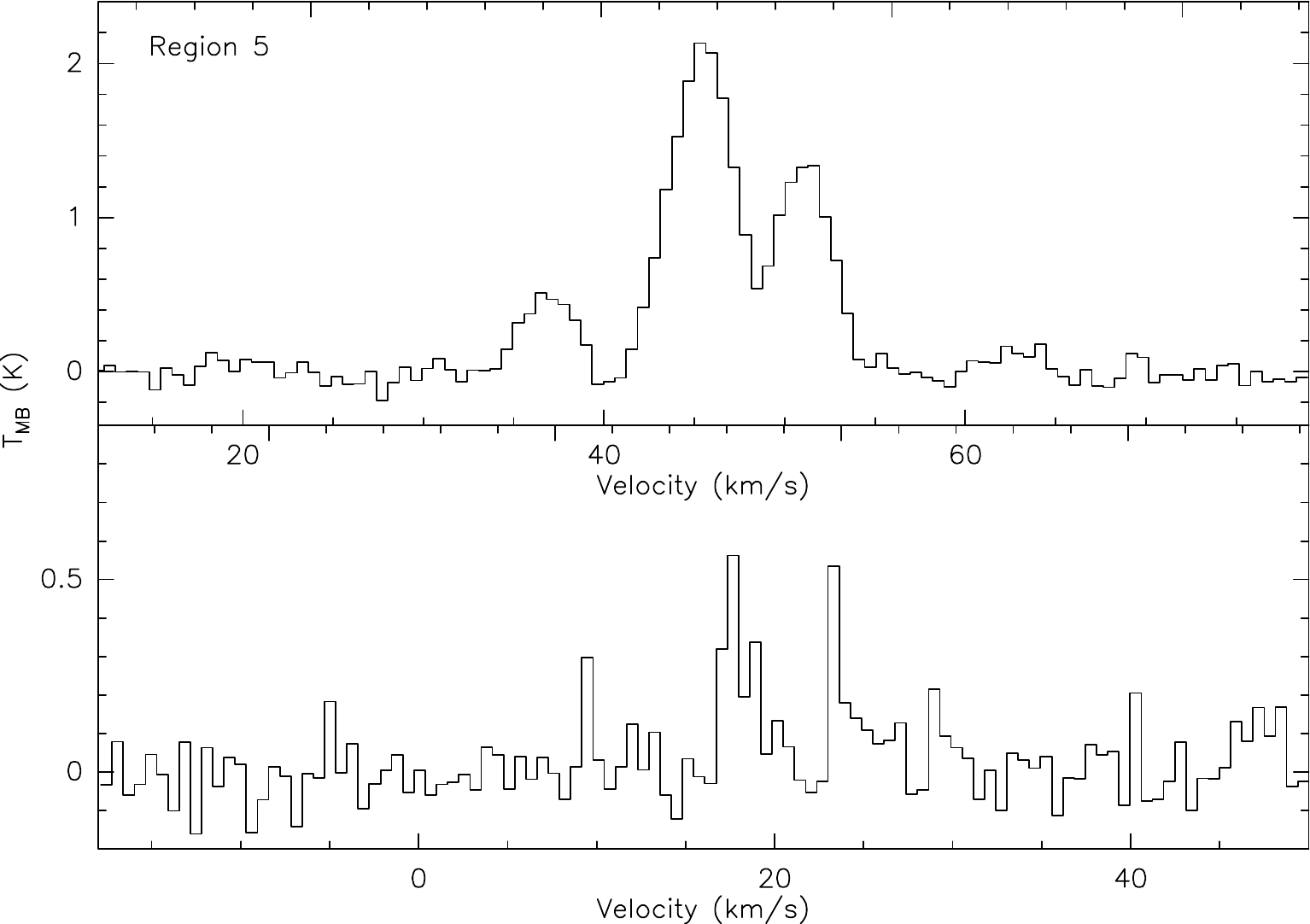}}
    \subfigure{\includegraphics[width=0.45\linewidth]{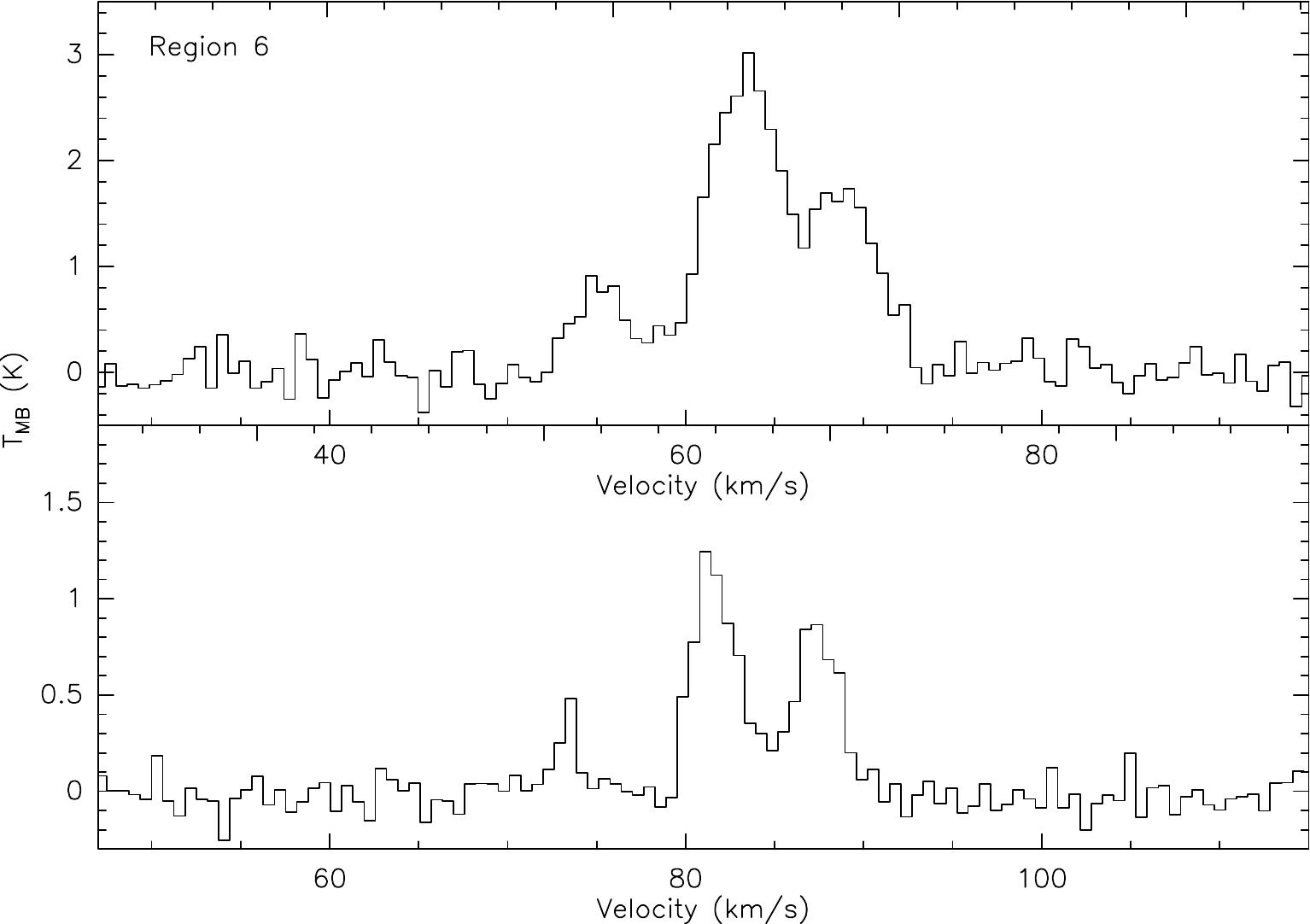}}
    \caption{Average spectra showing all velocity components of N$_2$H$^+$(1$-$0) found in the observed regions. The averaging was done centered at the integrated intensity peaks of the components and inside a single beam.}
    \label{sag_allspectra}
\end{figure*}

\begin{table*}
    \centering
    \footnotesize
    \begin{tabular}{l|l|l|l|l|l|l|l|l|l|l|l|l}
    ID & RA & DEC & $a$ & $b$ & PA & $v_{\rm LSR}$ & $\Delta v_{\rm HFS}$ & $\tau$ & $N$(N$_2$H$^+$) & $W$($^{13}$CO) & $W$(N$_2$H$^+$) & $N$(H$_2$)$_{\rm dust}$ \\
       & [$\degr$] & [$\degr$] & [$\arcsec$] & [$\arcsec$] & [$\degr$] & [km\,s$^{-1}$] & [km\,s$^{-1}$] &  & [10$^{14}$ cm$^{-2}$] & [K\,kms$^{-1}$] & [K\,kms$^{-1}$] & [10$^{22}$\,cm$^{-2}$] \\
       \hline \hline
       R1CL1  &  284.60697 & 3.03848 & 49.1 & 20.7 & -143 & 75.9 (0.2) & 1.77 (0.16) & 2.0 (0.8) & 0.27 (0.07)   & 15.4 (3.1) & 4.5 (0.9) & 2.00 (0.30)\\
       R1CL2  &  284.63488 & 3.04875 & 18.0 & 13.3 & 167  & 80.8* (0.0) & ... & ... & ...                        & 16.3 (2.8) & 2.6 (0.3) & 1.61 (0.06)\\ 
       R1CL3  &  284.59294 & 3.07755 & 24.5 & 17.1 & 60   & 75.0 (0.1)  & 1.25 (0.30) & 9.12 (1.4) & 0.48 (0.07)  & 6.2 (1.0) & 1.8 (0.2)  & 1.27 (0.03)\\ 
       R1CL4  &  284.61032 & 3.00986 & 17.4 & 13.1 & 105  & 79.5 (0.1) & 1.05 (0.0)   & 8.32* & 0.46*             & 3.9 (0.7) & 1.8 (0.3)  & 1.34 (0.02)\\ 
       R1CL5  &  284.59118 & 3.02467 & 19.8 & 17.8 & 146  & 88.6* (0.0) & ... & ... & ...                & 3.4 (0.7) & 2.5 (0.4)  & 1.72 (0.06)\\ 
       R1CL6  &  284.59325 & 3.04520 & 39.0 & 20.9 & 88   & 77.7 (1.6)  & 1.26 (0.24) & 3.81 (2.7) & 0.33 (0.19)& 4.4 (1.8) & 2.5 (0.4)  & 1.71 (0.10)\\ 
       R1CL7  &  284.60066 & 3.09832 & 23.6 & 17.2 & 117  & 52.5 (0.2) & 1.05 (0.0)   & 5.58 (2.8) & 0.50 (0.15)  & 2.3 (0.8) & 1.7 (0.2)  & 1.54 (0.07) \\
       R2CL1  &  284.76666 & 3.42842 & 15.9 & 13.4 & 123 & 78.5 (0.3) & 1.05 (0.00)  & ... & ...  & 15.2 (0.7) & 2.8 (0.4) & 1.68 (0.12)  \\
       R2CL2  &  284.76436 & 3.44348 & 21.5 & 18.5 & 117 & 83.9 (0.0) & ... & ...         & ...           & 5.3  (1.7) & 4.1 (2.2) & 1.24 (0.04) \\ 
       R2CL3  &  284.80803 & 3.34650 & 22.8 & 19.8 & 67  & 77.2 (0.1) & 1.37 (0.15) & ... & ...  & 7.1  (1.5) & 1.3 (1.5) & 1.95 (0.17) \\ 
       R2CL4  &  284.81486 & 3.34983 & 15.9 & 6.9 & -159 &  ...         & ... & ...         & ...           & 5.2  (0.0) & 3.2 (1.7) & 1.55 (0.07) \\
       R3CL1  &  284.76949 & 3.89697 & 56.7 & 40.1 & 178 & 91.0 (1.2) & 2.11 (0.13) & 1.3 (0.5) & 0.30 (0.11) & 13.1 (5.7) & 12.4 (5.0) & 3.72 (3.00) \\
       R3CL2  &  284.71041 & 3.81477 & 68.6 & 29.1 & 123 & 89.9 (1.1) & 1.70 (0.47) & 2.8 (1.8) & 0.20 (0.06) & 3.4 (1.5) & 4.8 (2.1) & 1.33 (0.12) \\
       R3CL3  &  284.71874 & 3.97830 & 33.3 & 17.8 & 78  & 92.1 (0.1) & 1.26 (0.13) & 3.0 (0.9) & 0.18 (0.05) & 8.9 (1.8) & 1.8 (0.7) & 1.85 (0.47) \\
       R3CL4  &  284.78565 & 3.94774 & 27.9 & 18.8 & -162 & 90.4 (0.1)& 2.00 (0.23) & 1.7 (0.4) & 0.15 (0.01) & 11.3 (3.5) & 5.2 (0.7) & 2.15 (0.56) \\
       R3CL5  &  284.60774 & 3.78785 & 27.8 & 23.5 & 90  &  81.5 (0.1)& 1.23 (0.19) & 3.0 (0.9) & 0.21 (0.06) & 8.3 (3.6) & 3.3 (1.0) & 1.23 (0.11) \\
       R4CL1  &  284.78465 & 4.26570 & 19.1 & 12.0 & 129 & 95.1* (...) & 1.05* & 4.6* & 0.24* &                           5.1 (1.5) &  3.2 (0.6) & 1.41 (0.03)  \\ 
       R4CL2  &  284.79428 & 4.20338 & 73.8 & 43.8 & 158 & 84.1 (0.3) & 2.45 (0.37) & 2.0 (0.9) & 0.21 (0.06) & 12.2 (7.7) & 12.0 (3.7) & 3.80 (4.10) \\
       R4CL3  &  284.89847 & 4.28015 & 28.5 & 14.8 & 95  & 86.7 (0.1) & 2.05 (0.26) & 8.2 (2.0) & 0.29 (0.02) & 17.6 (4.6) & 2.4 (0.6)  & 2.84 (0.28) \\
       R4CL4  &  284.03724 & 4.26913 & 24.3 & 13.9 & 160 & 86.3 (0.1) & 1.76 (0.18) & 5.4 (0.9)  & 0.19 (0.03) &        3.0 (1.0) &  3.3 (0.4)  & 1.26 (0.03) \\
       R4CL5  &  284.79507 & 4.27264 & 29.6 & 13.7 & 167 & 86.3 (0.3) & 1.19 (0.20) & 5.9 (0.7)  & 0.21 (0.01) &        2.3 (2.3) &  3.4 (0.8)  & 1.47 (0.02) \\
       R4CL6  &  284.91742 & 4.24616 & 58.6 & 25.4 & 118 & 82.9 (0.3) & 1.52 (0.33) & 5.6 (3.0)  & 0.19 (0.08) &        4.1 (2.2) &  3.7 (0.8)  & 1.72 (0.12) \\
       R4CL7  &  284.91916 & 4.29636 & 67.7 & 29.1 & 86  & 88.8 (2.7) & 1.64 (0.08) & 4.1* & 0.14* &                       3.1 (2.7) & 2.4 (1.3)   & 1.35 (0.07) \\
       R4CL8  &  284.83787 & 4.18411 & 25.2 & 19.6 & 116 & 40.8 (0.1) & 1.25 (0.10) & 2.3 (1.0) & 0.27 (0.11) & 4.0 (1.8) &  1.5 (0.2)  & 1.67 (0.13) \\
       R5CL1  &  285.02672 & 3.99503 & 44.1 & 21.6 & 52  & 56.9 (2.4) & 1.83 (0.21)  & 2.2 (1.1) & 0.19 (0.03) & 23.2 (6.2) & 5.4 (1.0) & 2.61 (0.60) \\
       R5CL2  &  284.93555 & 3.84254 & 14.6 & 9.0  & 125 & 58.1 (0.2) & ...          & 5.6 (1.8) & 0.18 (0.06) & 14.3 (1.1) & 1.3 (0.1) & 2.13 (0.26) \\
       R5CL3  &  284.96399 & 3.92188 & 16.7 & 10.6 & 176 & 56.6* (0.0) & 2.43* (0.75) & 3.9 (0.1) & 0.18 (0.03) & 16.7 (0.3) & 4.3 (0.5) & 2.26 (0.35) \\
       R5CL4  &  284.92858 & 3.89726 & 48.9 & 20.8 & 76  & 54.6 (0.3) & 1.79 (0.12)  & 2.2 (1.0) & 0.22 (0.05) & 22.1 (9.4) & 5.5 (2.5) & 3.27 (0.99) \\
       R5CL5  &  284.97721 & 3.99902 & 19.0 & 12.9 & -138 & 55.9 (0.2) & 1.84* & 18.2* & 0.62* & 19.6 (3.1) & 3.0 (0.5) & 2.09 (0.16) \\
       R5CL6  &  285.06609 & 4.05690 & 29.8 & 19.7 & 143 & 50.8 (0.2)  & 2.27 (0.30) & 2.0 (0.8) & 0.16 (0.04) & 39.1 (11.4) & 6.3 (1.1) & 4.49 (1.45) \\
       R5CL7  &  285.07437 & 4.16839 & 14.7 & 9.7  & 177 & 50.7 (3.7) & 1.34*  & 6.0* & 0.23* & 8.2 (0.0) &  1.7 (0.5) & 2.25 (0.10) \\
       R5CL8  &  285.15250 & 4.22302 & 47.2 & 31.0 & 165 & 45.3 (0.4) & 2.90 (0.27)  & 1.0 (0.5) & 0.22 (0.05) & 25.6 (13.3) & 14.4 (3.8) & 7.49 (5.37) \\
       R5CL9  &  285.09970 & 4.13984 & 16.2 & 10.6 & 131 & 45.6 (0.1) & 1.64 (0.06)  & 3.9 (0.5) & 0.16 (0.03) & 24.5 (2.0) & 2.7 (0.5) & 2.44 (0.18) \\
       R6CL1  &  285.25712 & 4.20239 & 54.6 & 38.9 & 157 & 63.7 (1.8) & 3.72 (0.25)  & 2.1 (1.9) & 0.48 (0.20) & 31.6 (7.9) & 19.1 (5.5) & 5.53 (2.72)\\
       R6CL2  &  285.43692 & 4.26914 & 20.0 & 17.3 & 76  & ...        & ...          & ...       & ...         & 24.0 (6.5) & 3.9 (0.9) & 4.43 (1.58) \\
       R6CL3  &  285.41506 & 4.16357 & 25.1 & 21.0 & -166 & 66.0 (1.9) & 2.35 (0.13) & 3.0 (1.0) & 0.20 (0.03) & 7.3 (2.3)  & 6.0 (0.9) & 3.93 (0.71)\\
       R6CL4  &  285.43425 & 4.17344 & 35.6 & 24.0 & 106 &  65.4 (0.1) & 2.42 (0.18) & 1.8 (0.7) & 0.17 (0.04) & 12.9 (3.0) & 6.4 (1.7) & 5.32 (1.56) \\
       R6CL5  &  285.23238 & 4.20370 & 34.4 & 20.8 & 134 &  61.2 (3.0) & 2.04* & 2.1* & 0.20* & 16.8 (6.5) & 6.5 (1.5) & 4.76 (2.05) \\
       R6CL6  &  285.48441 & 4.25207 & 49.4 & 31.5 & 131 & ...         & ...         & ...       & ...         & 8.4 (4.2)  & 1.4 (1.5) & 2.00 (0.26) \\
       R6CL7  &  285.47346 & 4.21256 & 28.9 & 23.9 & 106 & 63.3* (0.0) & ...         & ...       & ...         & 83.0 (21.0) & 5.9 (1.3) & 8.48 (4.67)
    \end{tabular}
    \caption{The parameters of the N$_2$H$^+$ clusters identified with the dendrogram method. (1) The identifyer of the cluster; (2,3) Equatorial coordinates; (4,5) Major and minor axis length; (6) Position angle from the equator in clockwise direction; (7,8) Average velocity and HF linewidth; (9) Average optical depth; (10) Average N$_2$H$^+$ column density; (11, 12) Average $^{13}$CO and N$_2$H$^+$ integrated intensity; (13) Average H$_2$ column density from dust emission. Missing values could not be computed due to insufficient S/N on the location of the cluster. The symbol * marks clusters where only a single value of a parameter was available (thus it is not an average).}
    \label{tab:region_clusters}
\end{table*}


\bsp	
\label{lastpage}
\end{document}